\documentstyle[prd,aps,amsmath,latexsym,amssymb,floats]{revtex}
\def \slas{\kern -6.2pt /}
\def \sla{\kern -5.4pt /}
\def \sl{\kern -4.0pt /}
\def \Cslas{\kern -6.8pt /}
\def \Dslas{\kern -7.4pt /}
\def \slass{\kern -7.4pt /}

\def \ii{{\mathrm{i}}}
\def \d{{\mathrm{d}}}

\def \pd{\partial}
\def \e{{\mathrm{e}}}

\def \lcx{\tilde{x}}
\def \tl#1{\overset{\kern 2pt\circ}{#1}}
\def \tll#1{\overset{\kern -1pt\circ}{#1}}
\def \TL#1{\overset{\kern -28pt \circ}{#1}}
\def \TLL#1{\overset{\kern -7pt \circ}{#1}}

\def \Tensor#1{\overset{\leftrightarrow}{#1}}

\newcommand{\be}{\begin{equation}}
\newcommand{\ee}{\end{equation}}
\newcommand{\bea}{\begin{eqnarray}}
\newcommand{\eea}{\end{eqnarray}}
\def \ba{\begin{align}}
\def \ea{\end{align}}

\begin{document}

\title{Power corrections of off-forward quark distributions\\ and
harmonic operators with definite geometric twist}
\author{Bodo Geyer$^1$\thanks{E-mail: geyer@itp.uni-leipzig.de},
	Markus Lazar$^1$\thanks{E-mail: lazar@itp.uni-leipzig.de},
	and 
	Dieter Robaschik$^2$\thanks{E-mail: drobasch@physik.tu-cottbus.de}}

\address{$^1$ Center for Theoretical Studies
	and Institute of Theoretical Physics,\\
	Leipzig University, 
	Augustusplatz~10, D-04109~Leipzig, Germany\\
	$^2$ BTU Cottbus, Fakult\"at 1, Postfach 101344, D-03013~Cottbus,
	Germany }

\date{\today}    
	
\maketitle
\hspace*{2cm}
\begin{abstract}
We introduce a group theoretically motivated procedure of parametrizing 
non-forward matrix elements of non-local QCD operators by (two-variable)
distribution amplitudes of well-defined geometric twist being multiplied 
by kinematical factors (related to the Lorentz structure of the operators 
and to the target states) as well as position-dependent coefficient
functions resulting from the (infinite) twist decomposition of the operators.
These distribution amplitudes are interpreted as (sum over) power corrections 
of the double distributions. --- Using the technique of harmonic
polynomials for the local operators we determine the (infinite) twist 
decomposition of totally symmetric operators completely and for operators 
with non-trivial symmetry type up to twist $\tau =3$. 
This covers the phenomenological interesting quark-antiquark operators.
Using these results we determine the power corrections to the various 
double distributions and the vector meson wave functions.
It is shown that the structure of the kinematical power corrections 
may be obtained, by harmonic extension, from the corresponding
expressions for operators or distribution amplitudes, on the light-cone. 
\hspace*{0.5cm}

\noindent
PACS number(s): 12.38 Bx, 13.85 Fb
\end{abstract}

\section{Introduction}
\setcounter{equation}{0}

The universal, non-perturbative parton distribution amplitudes 
parametrizing, modulo kinematical factors, the matrix elements of 
appropriate non-local quark-antiquark (as well as gluon) operators 
play a central role in phenomenological considerations. These 
operators occur in the quantum field theoretic description 
of light-cone dominated hadronic processes via the non-local light-cone 
expansion \cite{AZ78,ZAV,MRGHD}.
For deep inelastic lepton-hadron scattering and Drell-Yan processes 
the parton distributions are given as forward matrix elements of bilocal 
light-ray operators resulting from the time-ordered product of appropriate 
hadronic currents. For (deeply) virtual Compton scattering 
and hadron wave functions the so-called double distributions and hadron
distribution amplitudes, respectively, are given
by corresponding non-forward matrix elements.  Thereby, various processes 
are governed by one and the same (set of) light-ray operators.

For the sake of definiteness let us remember the amplitude of virtual 
Compton scattering, 
\begin{align}
\label{CA}
T_{\mu\nu}(P_i,Q; S_i) = \ii\int \d^4 x\, \e^{\ii qx}
\langle P_2,S_2| RT\left(J_\mu(x/2)J_\nu(-x/2){\cal S}\right)
| P_1,S_1 \rangle,
\end{align}
where $Q^2=-q^2,\,q= q_2-q_1$ denotes the momentum transfer and $\cal S$
is the (renormalized) $S-$matrix. Restricting to lowest order the 
renormalized time-ordered product of the electromagnetic (hadronic) 
currents, $J_\mu(x) = :\!\bar\psi(x) \gamma_\mu \psi(x)\!:$ 
in the vicinity of the light-cone, $x\rightarrow \lcx , \, \lcx^2 =0$,
where
\begin{align}
 \lcx = x - \zeta \Big((x\zeta) - \sqrt{(x\zeta)^2 - x^2}\Big)
\qquad {\rm with}\qquad
\zeta^2 = 1,
\end{align}
is approximated by (a sum over) coefficient functions 
$C^{(5)}_{\mu\nu\alpha}(x^2, \kappa\lcx,\mu^2)$ 
times the matrix elements of (renormalized) light-ray operators
($f$ denote the quark flavours)
\begin{align}
\label{OFC}
\langle P_2,S_2|O^{(5)}_\alpha(\kappa\lcx, -\kappa\lcx)|P_1,S_1\rangle
=
\sum_{n=0}^\infty\frac{(\ii\kappa)^n}{n!}
\langle P_2,S_2|
\bar{\psi}^f\!(0)(\gamma_5)\gamma_\alpha (\lcx\Tensor D)^n\psi^f\!(0)
|P_1,S_1\rangle.
\end{align}

Similar considerations hold for the (vector) meson wave functions
occurring in processes 
like exclusive semileptonic (or radiative) $B$ decays and hard 
electroproduction of vector mesons ($V=\rho$, $\omega$, $K^*$, $\phi$). 
The physical interest in this case arises from the fact that the following 
vertex function is directly observable in QCD:
\begin{align}
\label{TA}
T^V_{\mu\nu}(q,P;\lambda) = \int \d^4 x\, \e^{\ii qx}
\langle 0| RT\left(J_\mu(x/2)J_\nu(-x/2){\cal S}\right)
|V(P,\lambda) \rangle,
\end{align}
where $J_\mu$ can be axial or vector currents and $|V(P,\lambda)\rangle$
is the vector meson state of momentum $P$ and helicity $\lambda$.
Here, the universal nonperturbative quantities, the meson distribution 
amplitudes (DA), sometimes also called meson wave functions,
are given by means of the vacuum-to-meson matrix elements of bi-local
operators which, in the limit $x\rightarrow\lcx$, read
\begin{align}
\label{MFC}
\langle 0|O_\Gamma(\lcx,-\lcx) |V(P,\lambda)\rangle=
\sum_{n=0}^\infty\frac{\ii^n}{n!}
\langle 0|
\bar{\psi}^f(0)\Gamma (x\Tensor D)^n\psi^{f'}(0)|
V^{(f'f)}(P,\lambda)\rangle,
\end{align}
with $\Gamma = \{1, \gamma_5; \gamma_\alpha, \gamma_\alpha \gamma_5;
\sigma_{\alpha\beta}, \sigma_{\alpha\beta} \gamma_5 \}$.

Possibly, in the near future the experimental precision data  
will allow for the determination of
non-leading contributions and, therefore, require for a careful
analysis of the various sub-dominant effects contributing to the
physical processes. When considered beyond leading order, i.e.,~beyond
lowest twist operators in tree approximation, one is confronted 
not only with radiative corrections but also with power corrections
resulting from higher twist as well as target mass effects. 
Higher twist contributions are obtained by the decomposition of the
bilocal operators $O_\Gamma(\kappa x, -\kappa x)$
with respect to (irreducible) tensor representations of the Lorentz 
group having definite twist $\tau =$ (scale) 
dimension minus (Lorentz) spin \cite{GT71}. These representations are 
characterized
by their symmetry type (under index permutations of the traceless 
tensors) which is determined by corresponding Young frames having,
in the case of the Lorentz group, at 
the most three lines, $[\underline m] = [m_1,m_2,m_3]$.

The twist decomposition of non-local light-ray operators (`finite 
on-cone decomposition'), as far as they are relevant for hadronic 
processes, has been completed quite recently. A comprehensive presentation 
of these results can be found in Refs.~\cite{GLR99,GL00a} where also
earlier studies are mentioned. Recently it has been used for the 
definition of quark distributions functions of well-defined 
{\em geometric} twist together with their unique relationship to the 
quark distributions of {\em dynamical} twist (being introduced in 
Ref.~\cite{JJ91}). However, in order to get information
about the target mass contributions one is forced to consider the twist 
decomposition off-cone, taking into account all the trace terms leading 
to expressions suppressed by powers of $M^2/Q^2$.

For local operators which are characterized by symmetry type $[n]$,
i.e.,~being determined by totally symmetric traceless tensors, there exists 
a well-defined group theoretical framework \cite{BT77} which works
on the light-cone as well as, by unique harmonic extension, also off the
light-cone (`infinite off-cone decomposition'). 
For any other symmetry types which, however, are relevant
in the above mentioned physical applications no general theory exists.
But, from our earlier papers \cite{GLR99,GL00a} partial results can
be obtained. Using them we are able to present the decomposition of the
relevant off-cone operators up to twist $\tau = 3$ which already covers 
the cases being experimentally accessible in the future. In principle, 
twist-4 results also would be of interest, but there occur at least two 
bi-local operators of different symmetry types of which only one has been 
fully  determined.

Let us now explicitly list down, suppressing their flavour content and, 
as usual, the gauge link, the non-local quark operators 
which will be studied in detail. We determine the twist decomposition
of the following chiral-even (axial) vector operators
\begin{align}
\label{O_nl}
O_{\alpha}(\kappa x,-\kappa x)
&=\bar{\psi}(\kappa x)\gamma_{\alpha}\psi(-\kappa x),
\\
\label{O5_nl}
O_{5\alpha}(\kappa x,-\kappa x)
&=\bar{\psi}(\kappa x)\gamma_{\alpha}\gamma_5 \psi(-\kappa x),
\end{align}
{together with the corresponding (pseudo) scalar operators }
\begin{align}
\label{sO_nl}
O_{(5)}(\kappa x,-\kappa x)
&=x^\alpha O_{(5)\alpha}(\kappa x,-\kappa x),
\end{align}
{and the chiral-odd scalar and skew tensor operator}
\begin{align}
\label{N_nl}
N(\kappa x,-\kappa x)
&=\bar{\psi}(\kappa x) \psi(-\kappa x),\\
\label{M_nl}
M_{[\alpha\beta]}(\kappa x,-\kappa x)
&=\bar{\psi}(\kappa x)\sigma_{\alpha\beta}\psi(-\kappa x).
\end{align}
{together with the vector and scalar operators}
\begin{align}
\label{vM_nl}
M_{\alpha}(\kappa x,-\kappa x) 
&= x^\beta M_{[\alpha\beta]}(\kappa x,-\kappa x),
\\
\label{sM_nl}
M(\kappa x,-\kappa x) 
&= x^\beta \pd^\alpha M_{[\alpha\beta]}(\kappa x,-\kappa x).
\end{align}

The operators (\ref{O_nl}) -- (\ref{sM_nl}) constitute a basis not 
only for the (usual) parton
distributions as well as meson wave functions~\cite{GL01,L01a,L01b}
but also for the consideration of double distribution amplitudes  
being relevant for the various light-cone dominated QCD processes 
under consideration. Here, their twist decomposition
off the light-cone will be given up to twist 3 and, in the case of
scalar operators, also for any twist.
In fact, the `external' operation of contracting with $x^\alpha$ or 
$x^\beta\pd^\alpha$ also influences the possible symmetry type of these
operators and, therefore, of their twist decomposition. Notice that
the external coordinates are not multiplied by $\kappa$.

As a result these off-cone operators of definite twist are given 
for the local as well as the resummed nonlocal operators. The local 
operators given as the $n$-th Taylor coefficients of the non-local ones, 
see, Eqs.~(\ref{OFC}) and (\ref{MFC}) (or, more generally, 
Eqs.~(\ref{O^Gint}) and (\ref{O^Gloc}) below), are represented 
by (a finite series of) Gegenbauer polynomials $C^\nu_n(z),\, \nu \geq 1$. 
The nonlocal operators, being obtained by resummation
with respect to $n$, are represented by (a related series of) Bessel 
functions or, more exactly, either $J_{\nu-\frac{1}{2}}(z)$ or 
$I_{\nu-\frac{1}{2}}(z)$ depending on the values of their arguments. 

The group theoretical method for the determination of 
target mass corrections in unpolarized deep inelastic scattering
using harmonic scalar operators of definite spin and the corresponding 
matrix elements in terms of Gegenbauer polynomials has been used for
the first time by Nachtmann~\cite{Nachtmann73}. Some years later, 
using the same procedure,
the target mass contributions for polarized deep inelastic scattering 
were studied in Refs.~\cite{BE76,Wandzura77,MU80,KU95}. A short review of 
Nachtmann's method in deep inelastic lepton-hadron scattering was 
given in Ref.~\cite{GRW79}. 
This method differs from the one being applied by Georgi and 
Politzer~\cite{GP} for the target mass corrections in unpolarized
deep inelastic scattering. The latter method which is tailored to the 
forward case has been applied also to the study of target mass
corrections of polarized structure functions~\cite{PR98,BT99}.

Here, we generalize Nachtmann's procedure to the case of non-forward
matrix elements applying it to off-cone quark-antiquark operators.
In this paper we restrict ourselves to a fairly complete consideration of 
the various distribution amplitudes and postpone a more general application,
e.g., to the virtual Compton scattering (using a somewhat different
approach the latter has been already considered in Ref.~\cite{BM01}). 
However, we determine completely
the off-cone power corrections of the meson distribution amplitudes
in $x-$space which are much easier to handle.

The paper is organized as follows. 
In Sect.~II we present our method by considering a typical example.
The power of that approach consists in the fact that one determines
the twist decomposition at first for the local and nonlocal operators
and only afterwards takes the matrix elements.
In Sect.~III totally symmetric operators are studied thus obtaining 
the most general result including the whole series of infinite twist.
In Sect.~IV the twist decomposition of operators 
(\ref{O_nl}) -- (\ref{sM_nl}) up to twist 2 and 3 is given.
In Sect.~V these results are applied to the double distributions 
and the meson distributions.



\section{Non-forward matrix elements of 
nonlocal off-cone operators: The method}
\setcounter{equation}{0}

\subsection{Parametrization of non-forward matrix elements
by independent double distributions of definite twist}

To begin with we consider the bilocal off-cone quark-antiquark operators 
(\ref{O_nl}), (\ref{O5_nl}), (\ref{N_nl}) and (\ref{M_nl}),
i.e.,~operators {\em without} external operations which, generically,
will be denoted by ${\cal O}_{\Gamma}(\kappa x, -\kappa x)$. More
generally, they are given as
\begin{align}
{\cal O}_{\Gamma}(y+\kappa_1 x, y+\kappa_2 x)
= \bar\psi(y+\kappa_1 x) \Gamma 
U(y+\kappa_1 x, y+\kappa_2 x) \psi(y+\kappa_2 x),
\nonumber
\end{align} 
with $\eta = y+(\kappa_2+\kappa_1)(x/2)$ and 
$\xi = 2\kappa x\equiv (\kappa_2 - \kappa_1)x $ 
being the centre and the relative coordinate, respectively, 
\begin{align}
U(y+\kappa_1 x, y+\kappa_2 x) =
{\cal P}\exp \Big\{-\ii g\int^{\kappa_2}_{\kappa_1}\d t\,x^\mu 
A_\mu(y+tx)\Big\}
\nonumber
\end{align}
being the (straight) path ordered phase factor 
ensuring gauge invariance (which is omitted in the following).

The Fourier transforms of the `centred' operators 
${\cal O}_{\Gamma}(\kappa x, -\kappa x)$ and their $n$th moments
${\cal O}_{\Gamma n}(x)$, i.e., their Taylor coefficients
w.r.t.~$\kappa$, are related as 
follows:
\begin{align}
\label{O^Gint}
{\cal O}_{\Gamma}(\kappa x, -\kappa x)
=\int \d^4 q \,{\cal O}_{\Gamma}(q)\, \e^{\ii\kappa xq}
=\sum_{n=0}^\infty \frac{(\ii \kappa)^n}{n!}\,{\cal O}_{\Gamma n}(x),
\end{align}
{with}
\begin{align}
\label{O^Gloc}
{\cal O}_{\Gamma n}(x)
= \int \d^4 q \,{\cal O}_{\Gamma}(q)\, (xq)^n
= (-\ii)^n \frac{\pd^n}{\pd\kappa^n}
{\cal O}_{\Gamma}(\kappa x, -\kappa x)\Big|_{\kappa=0}.
\end{align}
For notational simplicity we wrote the Fourier
measure without the usual factor $1/(2\pi)^4$; obviously, 
$q$ should not be confused with some momentum transfer.
It should be remarked that the Taylor expansion of the non-local 
operators into local ones is only justified in a restricted
region of the Hilbert space \cite{AZ78,ZAV}. Here, we assume their 
existence after taking physical matrix elements.

Now, we introduce a suitable, but preliminary parametrization of the 
non-forward matrix 
elements of the bilocal operators ${\cal O}_{\Gamma}(\kappa x, -\kappa x)$. 
Namely, they can be represented as (everywhere, we use summation convention 
w.r.t. $a$)
\begin{align}
\label{NFME}
\langle P_2,S_2 |{\cal O}_{\Gamma}(\kappa x,-\kappa x)|P_1,S_1 \rangle
= {\cal K}_\Gamma^a({\mathbb P})
\int {\mathrm D}{\mathbb Z}\, \e^{\ii\kappa (x{\mathbb P}){\mathbb Z}}\,
f_a({\mathbb Z}, {\mathbb P}_i {\mathbb P}_j, x^2; \mu^2 ).
\end{align}
This parametrization which takes up and generalizes previous 
ones~\cite{MRGHD,J97,R97} deserves some explanations and further comments:
\begin{itemize}
\item{
First, we introduced the notation
${\mathbb P} = \{P_+, P_-\}$ and ${\mathbb Z} = 
\{z_+,  z_-\}$ with  $P_\pm = P_2\pm P_1$ and
$z_\pm=\hbox{\large$\frac{1}{2}$} (z_2 \pm z_1)$
thereby defining some (2-dimensional) vector space with scalar product
${\mathbb P}{\mathbb Z} \equiv \sum P_i z_i = P_+z_+ + P_-z_-$.
In addition, the integration measure is defined by
${\mathrm D}{\mathbb Z}= dz_1 dz_2 \theta(1-z_1)\theta(z_1+1)
\theta(1-z_2)\theta(z_2+1)$.}
\item{
Next, ${\cal K}_\Gamma^a({\mathbb P})$ denote the linear independent spin
structures being defined by the help of the (free) hadron wave functions 
and governed by the $\Gamma-$structure of the corresponding
nonlocal operator ${\cal O}_{\Gamma}$. For example, in the case
of the virtual Compton scattering, there are two independent 
spin structures, the Dirac and the Pauli structure, 
${\cal K}_\mu^1=\bar u(P_2,S_2) \gamma_\mu u(P_1,S_1)$ and 
${\cal K}_\mu^2=\bar u(P_2,S_2) \sigma_{\mu\nu} P^\nu_- u(P_1,S_1)/M$,
respectively.}
\item{ 
Furthermore, modulo these spin structures, the Fourier transforms 
$f_a({\mathbb Z}, {\mathbb P}_i {\mathbb P}_j, x^2; \mu^2)$
of the matrix element 
$\langle P_2,S_2 |{\cal O}_{\Gamma}(\kappa x,-\kappa x)|P_1,S_1 \rangle$
with respect to the independent variables $\kappa (x{\mathbb P})$
are the associated (universal, renormalized) two-variable distribution 
amplitudes 
as introduced in \cite{MRGHD} but here extended off the light-cone.
As it is obvious, also their 
dependence on ${\mathbb P}_i {\mathbb P}_j$, $x^2$ and the renormalization 
point $\mu^2$ has to be taken into account. }
\item{
The representation (\ref{NFME}) gives the most general expression for the 
non-forward matrix elements of ${\cal O}_{\Gamma}(\kappa x,-\kappa x)$.
An essential aspect of that approach consists in the support restriction, 
$-1\leq z_i \leq 1$, of the distribution amplitudes 
$f_a({\mathbb Z}, {\mathbb P}_i {\mathbb P}_j, x^2; \mu^2)$ since
the matrix elements can be shown to be entire analytic functions with
respect to $x P_i$ (for a detailed discussion, see, \cite{ZAV,MRGHD}).}
\item{
Obviously, the representation (\ref{NFME}) may be used also
for more involved matrix elements, like (scalar) meson production, by 
extending the final state to $\langle P_2,S_2; k_1|$ thereby enlarging
the $\mathbb Z-$ and the $\mathbb P$--space and extending the set of 
possible spin structures $K^a_\Gamma({\mathbb P})$ and related distribution 
amplitudes $f_a$ (see,~e.g.,~Ref.~\cite{BEGR}). It can be
specified also to vacuum-to-hadron transition amplitudes.}
\end{itemize}

Finally, let us comment on operators {\em with} external operations, 
i.e.~when the
(axial) vectors or the (skew) tensors are multiplied `externally' by 
$x^\beta$ and/or $\pd^\alpha$. In these cases the `external' vector 
$x^\beta$ or tensor $x^\beta\pd^\alpha$ is assumed {\em not} to be Fourier 
transformed and the external operations have to be applied onto both sides 
of Eq.~(\ref{NFME}). Thereby the tensor structure of these external  
operations matches with the tensor structure of $K^a_\Gamma$ which 
by itself are independent of the coordinates. Despite of this the
external operations heavily influence the possible symmetry type of
the local operators and their decomposition into irreducible tensor
representations of the Lorentz group. 

Now, let us take into account that the non-local operators
${\cal O}_{\Gamma}(\kappa x,-\kappa x)$, formally, are given by infinite 
series of operators of growing (geometric) twist $\tau$
(using summation convention also w.r.to $\Gamma$), 
\begin{align}
\label{TWIST}
{\cal O}_{\Gamma}(\kappa x,-\kappa x)
&= \sum_{\tau \geq \tau_{\rm min}} c^{(\tau)\Gamma'}_{~~\Gamma}\!(x)\,
{\cal O}^{(\tau)}_{\Gamma'}(\kappa x,-\kappa x),
\\
\intertext{with}
\label{twist}
{\cal O}^{(\tau)}_{\Gamma}(\kappa x,-\kappa x)
&= {\cal P}^{(\tau)\Gamma'}_{~~\Gamma}
{\cal O}_{\Gamma'}(\kappa x,-\kappa x),
\\
\label{proj}
\big({\cal P}^{(\tau)} \times
{\cal P}^{(\tau')}\big)^{~~\Gamma'}_{\Gamma}
&= \delta^{\tau \tau'}{\cal P}^{(\tau)\Gamma'}_{~\Gamma},
\end{align}
where ${\cal P}^{(\tau)}_{\Gamma\Gamma'}$ are well-defined projection
operators. Also here we have to add some comments in order to make
the content of Eqs.~(\ref{TWIST}) -- (\ref{proj}) more definite:
\begin{itemize}
\item{
The projection operators ${\cal P}^{(\tau)\Gamma'}_{~~\Gamma}\!(x,\pd_x)$
which immediately act on the undecomposed operators 
${\cal O}_{\Gamma}(\kappa x,-\kappa x)$
depend on the coordinates and their derivatives and, eventually, contain 
additional integrations with respect to some auxiliary variables. Usually,
the summation over $\Gamma'$ is restricted to the same tensorial 
type. The specific form of the projections may be read off 
from the explicit twist decompositions, cf.,~e.g.,~Sec.~III, where various 
infinite twist series have been derived, as well as Sec.~IV,
Eqs.~(\ref{OPROJ}) -- (\ref{MPROJ}).}
\item{
The coefficient functions $c^{(\tau)\Gamma'}_{~~\Gamma}\!(x)$ essentially 
depend on (powers of) $x^2$ but, in principle, according to the
tensor structure of the operators they also could depend on $x^\mu$,
cf.,~e.g.,~Eqs.~(\ref{O_v_sym}). In addition, we remark that different
operators of the same twist may occur having, of course, different
coefficient functions.} 
\item{
Approaching the light-cone, $x\rightarrow\lcx$, the 
series (\ref{TWIST}) terminates at some finite value $\tau_{\rm max}$
since almost all of the coefficient functions vanish for $x^2=0$.
That situation has been considered extensively, e.g.,~in 
Refs.~\cite{GLR99,GL00a}}
\item{
Obviously, analogous decompositions exist for the local 
operators ${\cal O}_{\Gamma n}(x)$ which are uniquely related to
Eqs.~(\ref{TWIST}) -- (\ref{proj}). They have been derived in
parallel to the non-local operators in Secs.~III and IV.}
\item{
Finally, we remark that the relations (\ref{O^Gint}) and (\ref{O^Gloc}) 
between the local and the non-local operators,  
${\cal O}_{\Gamma n}(x)$ and ${\cal O}_{\Gamma}(\kappa x,-\kappa x)$,
and the Fourier transforms, ${\cal O}_{\Gamma}(q)$, also hold
for the corresponding operators of definite twist $\tau$.}
\end{itemize}
Introducing the decomposition (\ref{TWIST}) into the representation 
(\ref{NFME}) allows for an analogous decomposition of the distribution
amplitudes, $f_a({\mathbb Z}, {\mathbb P}_i {\mathbb P}_j, x^2; \mu^2 )$,
\begin{align}
\label{DDT0}
{\cal K}_{\Gamma}^a({\mathbb P})\,
f_a({\mathbb Z}, {\mathbb P}_i {\mathbb P}_j, x^2; \mu^2 )
&= \sum_{\tau \geq \tau_{\rm min}} c^{(\tau)\Gamma'}_{~~\Gamma}\!(x)\,
{\cal K}_{\Gamma'}^a({\mathbb P})\,
f^{(\tau)}_a({\mathbb Z}, {\mathbb P}_i {\mathbb P}_j, x^2; \mu^2 ) .
\end{align}
Then, for the matrix elements of operators ${\cal O}^{(\tau)}_{\Gamma}$ 
with definite geometric twist $\tau$ we obtain two different versions,
\begin{align}
\label{NFMET}
\langle P_2,S_2|
{\cal O}^{(\tau)}_{\Gamma}(\kappa x,-\kappa x)|P_1,S_1\rangle
&=
{\cal K}_{\Gamma}^a({\mathbb P})
\int {\mathrm D}{\mathbb Z}\; \e^{\ii\kappa (x{\mathbb P}){\mathbb Z}}\,
f^{(\tau)}_a({\mathbb Z}, {\mathbb P}_i {\mathbb P}_j, x^2; \mu^2 )
\nonumber
\\
&= {\cal P}^{(\tau)\Gamma'}_{~~\Gamma}\!(x,\pd_x) \,
{\cal K}_{\Gamma'}^a({\mathbb P})
\int {\mathrm D}{\mathbb Z}\; \e^{\ii\kappa (x{\mathbb P}){\mathbb Z}}\,
f^{(\tau)}_a({\mathbb Z}; \mu^2 ),
\end{align}
where the first line is only the reduction of relation (\ref{NFME}) to 
their twist components and 
where, in the second line, using properties (\ref{proj}) of the 
projection (\ref{twist}) onto the non-local operators of definite twist 
we introduced the related Lorentz invariant {\em double distributions} 
$f^{(\tau)}_a({\mathbb Z};\mu^2)$ of definite twist $\tau$ not suffering
from any power corrections.
Concerning that definition of double distributions let us give a heuristic
justification and some of its consequences:
\begin{itemize}
\item{
Reminding the Fourier representation (\ref{O^Gint}) for the non-local
operators $O^{(\tau)}_\Gamma(\kappa x, -\kappa x)$ it looks reasonable
that the $x^2-$ and $\mathbb P-$dependence of the distribution amplitudes 
$f_a({\mathbb Z},{\mathbb P}_i{\mathbb P}_j,x^2;\mu^2 )$ should be already 
uniquely determined by the projection operators of (\ref{twist}) acting on 
the exponential on the right hand side of (\ref{NFMET}) and, possibly, on 
some external operations matching ${\cal K}_{\Gamma'}^a({\mathbb P})$.
This reflects the fact that every local operator of definite twist is 
determined by its symmetry type expressed, within the polynomial method,  
through specific differential operators acting on (traceless) harmonic tensor
operators containing the $x^2-$dependence, cf.,~Eq.~(\ref{equiv}) below.}
\item{
As a consequence, the mass corrections of the matrix elements (\ref{NFME}) 
are determined by Eq.~(\ref{NFMET}). Likewise, the distribution amplitudes  
$f_a({\mathbb Z}, {\mathbb P}_i {\mathbb P}_j, x^2; \mu^2 )$, using both 
forms of the matrix element (\ref{NFMET}), formally may be expressed
through the double distributions $f^{(\tau)}_{a'}({\mathbb Z}; \mu^2 )$
according to: 
\begin{align}
\label{DDT}
f^{(\tau)}_a({\mathbb Z}, {\mathbb P}_i {\mathbb P}_j, x^2; \mu^2 )
&= 
\big({\cal K}^{-1}({\mathbb P})\big)^{\Gamma}_a
\bigg(\e^{-\ii\kappa (x{\mathbb P}){\mathbb Z}}\,
{\cal P}^{(\tau)\Gamma'}_{~\Gamma}(x,\pd_x) 
\, \e^{\ii\kappa (x{\mathbb P}){\mathbb Z}}\bigg)\,
{\cal K}_{\Gamma'}^{a'}({\mathbb P})\,
f^{(\tau)}_{a'}({\mathbb Z}; \mu^2 )
\nonumber
\\
&=
{\cal F}^{(\tau)}_{aa'}\big((x{\mathbb PZ}), x^2 ({\mathbb PZ})^2\big)\,
f^{(\tau)}_{a'}({\mathbb Z}; \mu^2 );
\end{align}
here, 
${\cal F}^{(\tau)}_{aa'}\big((x{\mathbb PZ}),({\mathbb PZ})^2 x^2\big)$
contains any information about the power corrections of the double 
distribution. In addition, we observe that, as a result of the twist 
projection, the $(x\mathbb P)-$dependence of the matrix element is only
partly determined by the double distributions 
$f^{(\tau)}_{a'}({\mathbb Z}; \mu^2 )$.}
\item{
Obviously, since the decomposition of the distribution amplitudes is 
uniquely related to the decomposition of the corresponding 
(quark-antiquark) operators the power corrections of the double 
distributions, 
Eq.~(\ref{DDT}), are already determined by the {\em twist decomposition of} 
these {\em operators}. Both decompositions, besides on $(x\mathbb PZ)$, also 
depend on the following combination of variables,
$
 (x{\mathbb PZ})/ \sqrt{x^2 ({\mathbb PZ})^2}
{\rm ~and~} 
 \sqrt{(x{\mathbb PZ})^2 - x^2 ({\mathbb PZ})^2},
$ 
being the arguments of appropriate Gegenbauer polynomials and Bessel
functions in the case of local and non-local operators, respectively;
cf.,~Subsecs.~II~B and II~C below.}
\item{
Notice, that the double 
distributions of definite twist $f^{(\tau)}_a({\mathbb Z};\mu^2)$
being independent on $x^2$ are already determined by the decomposition of
the non-local matrix elements on the light-cone. Also their renormalization
properties are determined by the light-cone operators only.
}
\end{itemize}

This procedure of introducing double distributions of definite geometric
twist generalizes to arbitrary off-cone values of $x$ what has been
introduced in \cite{GL01} for the parton distribution functions of definite
geometric twist, i.e.,~for the case of forward matrix elements restricted
to the light-cone, and later on applied to the (vector) meson distribution
amplitudes \cite{L01a}, i.e.,~to vacuum-to-meson matrix elements, also 
restricted to the light-cone. 
\medskip

Let us now present the arguments leading to the r.h.s.~of Eq.~(\ref{NFMET}) 
more explicitly. Namely, the non-forward matrix elements after performing the 
twist projection of relations (\ref{O^Gint}) are given as follows:
\begin{align}
\label{equiv}
\langle P_2,S_2|{\cal O}^{(\tau)}_{\Gamma}(\kappa x,-\kappa x)|P_1,S_1 \rangle
&=
\langle P_2,S_2 |{\cal P}^{(\tau)\Gamma'}_{~~\Gamma}(x,\pd)\int \d^4q 
\,{\cal O}_{\Gamma'}(q)\,\e^{\ii\kappa xq} | P_1,S_1 \rangle
\nonumber\\
&=
\int \d^4q\,
\Big({\cal P}^{(\tau)\Gamma'}_{~~\Gamma}(x,\pd)\,\e^{\ii\kappa xq}\Big)
\langle P_2,S_2 |{\cal O}_{\Gamma'}(q) | P_1,S_1 \rangle
\\
&= 
\int {\mathrm D}{\mathbb Z}
\Big({\cal P}^{(\tau)\Gamma'}_{~~\Gamma}\!(x,\pd_x) \,
\e^{\ii\kappa (x{\mathbb P}){\mathbb Z}}\Big) \,
{\cal K}_{\Gamma'}^a({\mathbb P})\,f^{(\tau)}_a({\mathbb Z}; \mu^2 ).
\nonumber
\end{align}
Here, in the second line we simply reordered the operations of integration,
taking matrix elements and twist projections appropriately. In order to get 
the third line we observe the symmetry in $x$ and $q$ of the local twist 
projections,
${\cal P}^{(\tau)\Gamma'}_{n~\Gamma}(x,\pd) (xq)^n \equiv
{\cal P}^{(\tau)\Gamma'}_{n~\Gamma}(q,\pd_q) (xq)^n$, 
from which one derives for the second line:
\begin{align}
\sum_{n=0}^\infty \frac{(\ii\kappa)^n}{n!}
\Big({\cal P}^{(\tau)\Gamma'}_{n~\Gamma}(x,\pd)\,
x^{\mu_1} \ldots x^{\mu_n}\Big)
\int \d^4q\,
\Big({\cal P}^{(\tau)\Gamma''}_{n~\Gamma'}(q,\pd_q)\,
q_{\mu_1} \ldots q_{\mu_n}\Big)
\langle P_2,S_2 |{\cal O}_{\Gamma'' n}(q)| P_1,S_1 \rangle.
\nonumber
\end{align}
Here, the integrand contains the matrix elements of the irreducible local 
operators of definite twist ${\cal O}^{(\tau)}_{\Gamma\mu_1\ldots\mu_n}(q)$.
Their reduced matrix elements $f^{(\tau)}_{n_1n_2},~n_1+n_2=n,$ which are 
related to the decomposition of $(P_2+P_1)^n$ into independent monomials 
may be represented by (double) moments of corresponding double distributions 
$f_a^{(\tau)}(z_1,z_2)$ multiplied by the kinematical factors 
${\cal K}_\Gamma^a(P_1,P_2)$. Finally, since $q$ reflects the dependence
of the (local) operators on $\Tensor D$, after performing the integration
and the resummation over $n$, 
$q$ within the integrand simply has to be replaced by $\mathbb PZ$.
That procedure will be demonstrated explicitly in Subsec.~II.C.

Formally, in the various experimental situations, in any Fourier integrand 
having the structure of Eq.~(\ref{equiv}), we only have to perform the 
following replacements (denoted by $\doteq$):
\smallskip

\noindent
(A) In the case of {\em non-forward scattering}, as just explained, we obtain 
\begin{align}
\label{NLME}
\langle P_2,S_2 |{\cal O}_{\Gamma}(q) | P_1,S_1 \rangle
\doteq
{\cal K}_\Gamma^a({\mathbb P})
\int {\mathrm D}{\mathbb Z}\, \delta^{(4)}(q-{\mathbb P}{\mathbb Z})\,
f^{(\tau)}_a({\mathbb Z};\mu^2). 
\end{align}

\noindent
(B) In the case of {\em forward scattering}, i.e.,~for $P_1 = P_2 = P$, 
the situation changes into  
\begin{align}
\langle P,S |{\cal O}_{\Gamma}(q) | P,S \rangle
&\doteq
{\cal K}_\Gamma^{a}(P)
\int \d z\, \delta^{(4)}(q-2Pz)\,\hat f^{(\tau)}_a(z; \mu^2),
\end{align}
with $P_+ = 2P, P_- = 0; z_+ = z$. These 
distributions $\hat f^{(\tau)}_a(z,\mu^2)$  are obtained 
from the double distributions $\hat f^{(\tau)}_a({\mathbb Z},\mu^2)$
by integrating out the independent variable $z_-$, i.e.,
\begin{align}
\hat f^{(\tau)}_a(z,\mu^2)= \int \d z_- f^{(\tau)}_a(z_+=z, z_-; \mu^2).
\end{align} 

\noindent
(C) In the case of {\em vacuum-to-hadron transition amplitudes}, 
e.g.,~for the meson distribution amplitudes one obtains
\begin{align}
\label{NLM}
\langle 0 |{\cal O}_{\Gamma}(q) | V(P,\lambda) \rangle
&\doteq
\widetilde{\cal K}_\Gamma^{a}(P)
\int \d \xi\, \delta^{(4)}(q-P\xi)\,\tilde f^{(\tau)}_a(\xi;\mu^2),
\end{align}
with $P_2=P; z_2 = \xi$. Obviously, the spin structures are different
from the above cases and, by construction, only $P$ and $\xi$ occur.

From the foregoing discussion it is obvious that the mass corrections to the 
various physical processes are completely determined by the twist structure 
of the bilocal operators ${\cal O}_{\Gamma}^{(\tau)}(\kappa x, -\kappa x)
=\int\d^4 q\,{\cal O}_{\Gamma}^{(\tau)}(q)\,\e^{\ii\kappa xq}$. Performing 
matrix elements according to the replacements Eqs.~(\ref{NLME}) -- 
(\ref{NLM}) we obtain the related expressions for 
the $S$-matrix elements of the physical process under consideration.
Then, a further Fourier transformation with respect to the coordinate $x$ 
like in Eqs.~(\ref{CA}) and (\ref{TA}), finally, leads 
to a representation of the scattering or transition amplitudes in terms 
of the inverse momentum transfer depending, quite generally, on
$P_iP_j/Q^2$.

In the next two subsections we demonstrate our procedure for the simplest 
nontrivial case, first, directly for non-forward matrix elements and then, 
more efficiently, for the corresponding nonlocal operator itself. 
Thereby, we introduce the technique of 
expansion into (local) harmonic polynomials and their resummation to 
non-local harmonic operators.



\subsection{Non-forward matrix elements of bilocal operators 
having definite twist: A simple nontrivial example}
Let us now demonstrate our procedure by the simplest nontrivial example, 
the non-forward matrix element of the twist-2 part of the (pseudo)scalar 
operator, Eq.~(\ref{sO_nl}), which is relevant for the leading terms of 
virtual Compton scattering~\cite{BGR99}. 
This operator containing the `external' truncation by $x^\mu$ 
is given by \cite{BB88,GLR99}
\begin{align}
\label{OP2}
O^{\rm tw2}(\kappa x, -\kappa x) 
=&\;
\bar\psi(\kappa x) (x\gamma) \psi(-\kappa x) 
+
\sum_{k=1}^\infty
\int_0^1 \d t\, \Big(\frac{1-t}{t}\Big)^{k-1} \Big(\frac{-x^2}{4}\Big)^{\!k}
\frac{\square^k}{k!(k-1)!}
\,
\bar\psi(\kappa tx) (x\gamma) \psi(-\kappa tx). 
\end{align}
Its matrix elements, taking into account the Dirac structure only, read
\begin{align}
\label{nfme}
\langle P_2 |O^{\rm tw2}(\kappa x, -\kappa x) | P_1 \rangle
=&\;
\bar u(P_2) (x\gamma) u(P_1) 
\int {\mathrm D}{\mathbb Z}\, \e^{\ii\kappa (x{\mathbb P}){\mathbb Z}}\,
F_D^{(2)}({\mathbb Z})
\nonumber\\
&+
\sum_{k=1}^\infty
\int_0^1 \d t \Big(\frac{1-t}{t}\Big)^{k-1} \Big(\frac{-x^2}{4}\Big)^{\!k}
\frac{\square^k}{k!(k-1)!}
\,\bar u(P_2) (x\gamma) u(P_1)
\int {\mathrm D}{\mathbb Z}\, \e^{\ii t\kappa (x{\mathbb P}){\mathbb Z}}\,
F_D^{(2)}({\mathbb Z}),
\end{align}
where $F_D^{(2)}({\mathbb Z})$ denotes the twist-2 Dirac type double
distribution. Obviously, on the light-cone only the first term survives.
Here, we take the complete expansion into account.
Performing the differentiations, 
\begin{align}
\square^k (x\gamma)\e^{\ii\kappa t x{\mathbb P}{\mathbb Z}}
=
\Big[t^{2k}\,(x\gamma)\, (i\kappa x{\mathbb P}{\mathbb Z})^{2k}
+
2k\,t^{2k-1}\,(i\kappa \gamma{\mathbb P}{\mathbb Z})
(i\kappa x{\mathbb P}{\mathbb Z})^{2(k-1)}\Big]\,
\e^{\ii\kappa t x{\mathbb P}{\mathbb Z}} ,
\nonumber
\end{align}
and using
\begin{align}
\int_0^1 \d t\, \frac{(1-t)^{k-1}t^{\ell+1}}{(k-1)!\ell!}\,
\e^{\ii t\kappa (x{\mathbb P}){\mathbb Z}}
=
\sum_{n=0}^\infty \frac{(\ii\kappa x{\mathbb PZ})^n}{n!}
\frac{(n+\ell+1)!}{(n+k+\ell+1)! \ell!},
\nonumber
\end{align}
this can be rewritten as follows:
\begin{align}
\langle P_2 |O^{\rm tw2}(\kappa x, -\kappa x) | P_1 \rangle
=
&\int {\mathrm D}{\mathbb Z}\, F_D^{(2)}({\mathbb Z})
\bigg\{\bar u(P_2) (x\gamma) u(P_1) 
\sum_{k=0}^\infty
\sum_{n=0}^\infty \frac{(\ii\kappa x{\mathbb PZ})^n}{n!}
\frac{(n+k+1)!}{(n+2k+1)! k!}
\Big(\frac{(\kappa x)^2({\mathbb PZ})^2}{4}\Big)^{\!k}
\nonumber\\
&\quad
-\hbox{\large$\frac{1}{2}$}
\ii\kappa x^2\bar u(P_2) (\gamma{\mathbb P}{\mathbb Z}) u(P_1) 
\sum_{k-1=0}^\infty
\sum_{n=0}^\infty \frac{(\ii\kappa x{\mathbb PZ})^n}{n!}
\frac{(n+k)!}{(n+2k)! (k-1)!}
\Big(\frac{(\kappa x)^2({\mathbb PZ})^2}{4}\Big)^{\!k-1}
\bigg\}.
\nonumber
\end{align}
Now, shifting $k \rightarrow k+1$ in the second $k-$summation, and
observing the series representation of the modified Bessel functions
of the first kind (cf.,~Ref.~\cite{PBM}, Eq.~I.5.2.13.27),
\begin{align}
\sum_{n=0}^\infty 
\frac{\Gamma(n+\nu+\frac{1}{2})}{\Gamma(n+2\nu+1)n!}\, z^n
= \sqrt{\pi}\,  \e^{ z/2}\, z^{-\nu}\, I_\nu(z/2)
\end{align}
for $\nu = k+3/2$ and $z=i\kappa (x{\mathbb P}){\mathbb Z}$ we arrive at
\begin{align}
\langle P_2 |O^{\rm tw2}(\kappa x, -\kappa x) | P_1 \rangle
=
&\int {\mathrm D}{\mathbb Z}\, F_D^{(2)}({\mathbb Z})
\Big\{
\bar u(P_2) (x\gamma) u(P_1)(2+x\pd) 
- \hbox{\large$\frac{1}{2}$}
\ii\kappa x^2\bar u(P_2)(\gamma{\mathbb PZ})u(P_1)
\Big\}(3+x\pd)
\nonumber\\
&\qquad\qquad\times
\sum_{k=0}^\infty
\frac{\big((\kappa x)^2({\mathbb PZ})^2/4\big)^k}{k!}
\sum_{n=0}^\infty 
\frac{(n+k+1)!}{(n+2k+3)!}
\frac{(\ii\kappa x{\mathbb PZ})^n}{n!}
\nonumber\\
=
&\int {\mathrm D}{\mathbb Z}\, F_D^{(2)}({\mathbb Z})\,
\bar u(P_2) \gamma^\mu u(P_1)
\Big\{x_\mu (2+x\pd)
 - 
\hbox{\large$\frac{1}{2}$}
\ii\kappa\, {\mathbb P}_\mu{\mathbb Z}\, x^2
\Big\}(3+x\pd)
\nonumber\\
&\qquad\qquad\times
\sum_{k=0}^\infty 
\frac{\big((\kappa x)^2({\mathbb PZ})^2/4\big)^k}{k!}\,
\sqrt{\pi}\, \e^{\ii\kappa x{\mathbb PZ}/2} \,
(\ii\kappa x{\mathbb PZ})^{-k-3/2}\,
I_{k+3/2}(\ii\kappa x{\mathbb PZ}/2).
\nonumber
\end{align}
Now, using $I_\nu\big(\e^{\ii\pi/2} z) = \e^{\ii\pi\nu/2}\,J_\nu(z)$
with the (usual) Bessel functions $J_\nu(z)$ 
as well as (cf.~Ref.~\cite{PBM}, Eq.~II.5.7.6.1)
\begin{align}
\sum_{k=0}^\infty \frac{t^k}{k!} J_{k+\nu}(z)
= z^{\nu/2}\, (z-2t)^{-\nu/2} J_\nu\big(\sqrt{z^2-2tz}\big)
\qquad
{\rm for}
\qquad
|2t| < z,
\end{align}
and substituting $t = \kappa x^2 q^2 / (4 xq)$ and $z=\kappa xq/2$
we finally arrive at
\begin{align}
\langle P_2 |O^{\rm tw2}(\kappa x, -\kappa x) | P_1 \rangle
=
&\sqrt{\pi} \,
\bar u(P_2) \gamma^\mu u(P_1)
\int {\mathrm D}{\mathbb Z}\, F_D^{(2)}({\mathbb Z})\,
\Big\{x_\mu\, (2+x\pd)
- \hbox{\large$\frac{1}{2}$}
\ii\kappa \,{\mathbb P}_\mu{\mathbb Z}\, x^2\Big\}(3+x\pd)
\nonumber\\
\label{NLO2}
&\qquad\qquad\times
\Big(\kappa 
\sqrt{(x{\mathbb PZ})^2- x^2({\mathbb PZ})^2}
\Big)^{-3/2}
J_{3/2}\Big(\hbox{\large$\frac{\kappa}{2}$}
\sqrt{(x{\mathbb PZ})^2- x^2({\mathbb PZ})^2}\Big)
\,\e^{\ii\kappa x{\mathbb PZ}/2} .
\end{align}
This holds in the case $(x{\mathbb PZ})^2-x^2 {\mathbb PZ}^2 \geq 0$, 
otherwise we have to change into $I_\nu(z)$.
That result exactly corresponds to Eq.~(\ref{NFMET}) for the example
under consideration. Observe that another half of the exponential is
concealed within the Bessel function!

In the case of {\em forward scattering} 
because of $\bar u(P) \gamma^\mu u(P) = 2 P^\mu$ we obtain
\begin{align}
\langle P |O^{\rm tw2}(\kappa x, -\kappa x) | P \rangle
=
&\,2\sqrt{\pi} \,\int \d z\, \widehat F^{(2)}_D(z)\,
\Big\{(xP)\, (2+x\pd)- \ii\kappa z\,P^2 x^2\Big\}(3+x\pd)
\nonumber\\
\label{SNLO2}
&\qquad\qquad\times
\Big(2\kappa z
\sqrt{(xP)^2- x^2 P^2}
\Big)^{-3/2}
J_{3/2}\Big(\kappa z\sqrt{(xP)^2- x^2 P^2}\Big)
\,\e^{\ii\kappa xPz}.
\end{align}

If this expression is restricted to the light-cone, $x^2=0$, the
well-known parton distribution is obtained. Namely, using the Poisson
integral for the Bessel functions (cf.,~Ref.~\cite{BE},~Eq.~II.7.12.7),
\begin{align}
\label{Poisson}
\Gamma(\nu + \hbox{\large$\frac{1}{2}$}) J_\nu(z) 
= \frac{1}{\sqrt\pi} \Big(\frac{z}{2}\Big)^\nu
\int^1_{-1} \d t\,(1-t^2)^{\nu-1/2}\, \e^{\ii tz}
\qquad {\rm for} \qquad
{\rm Re~} \nu > - \hbox{\large$\frac{1}{2}$},
\end{align}
we obtain ($\xi = \kappa z (\lcx P)$)
\begin{align}
\langle P |O^{\rm tw2}(\kappa \lcx, -\kappa \lcx) | P \rangle
&=
2 (\lcx P) \sqrt{\pi} 
\int \d z\, \widehat F^{(2)}_D(z)\,(2+\xi\pd_\xi)(3+\xi\pd_\xi)
\big(2\xi \big)^{-3/2}
J_{3/2}(\xi)\,\e^{\ii\xi}
\nonumber\\
&=
2(\lcx P) \int \d z\, \widehat F^{(2)}_D(z)\,
\int^1_0 \d t\, t (1-t) (3+t\pd_t)(2+t\pd_t)\,\e^{2 \ii t \xi}
\nonumber\\
\label{lcNLO2}
&=
2(\lcx P) \int \d z\, \widehat F^{(2)}_D(z)\,
\int^1_0 \d t\, t (1+t\pd_t)\,\e^{2 \ii t \xi}
=
2 (\lcx P) \int \d z\,  F^{(2)}_D(z)\,\e^{ 2 \ii \kappa z (\lcx P)};
\end{align}
in the second line we introduced the Poisson integral after shifting the 
integration variable $t \rightarrow (t+1)/2$ and followed by changing
$\xi \pd_\xi  \rightarrow t\pd_t$, then we partially integrated two times
retaining finally only the surface term at $t=1$. As a result we arrived 
at the twist-2 parton distribution as introduced in  \cite{GL01},
$\widehat F^{(2)}_D(z) \equiv F^{(2)}(z)$, which coincides with
$f_1(z)$ in the notation adopted by Jaffe and Ji \cite{JJ91}.

The power corrections of the parton distribution $F^{(2)}(z)$ are
obtained according to the relation (\ref{DDT}). Namely, using the 
matrix element (\ref{SNLO2}) and, again, taking into account 
the representation (\ref{Poisson}), one gets
($Y=\sqrt{1- x^2 P^2/(xP)^2}$):
\begin{align}
\label{PD2}
F^{(2)}(z, xP, x^2P^2; \mu^2)
&=
F^{(2)}(z)\,\e^{-2\ii\kappa (xP)z}
\frac{1}{8}
\Big\{(2+x\pd)- \ii\kappa z\,\frac{x^2 P^2}{(xP)}\Big\}(3+x\pd)
\!\int^1_{-1}\!\! \d t\, (1-t^2) \,
\e^{\ii\kappa\,t\,(xP)z\,Y}
\,\e^{\ii\kappa (xP)z}\!
\nonumber\\
&=
F^{(2)}(z) \,\e^{-2\ii\kappa(xP)z}\,\frac{1}{4Y}
\Big\{
\big(1+Y\big)^2\e^{\ii\kappa(xP)z(1+Y)}
-
\big(1-Y\big)^2\e^{\ii\kappa(xP)z(1-Y)}
\Big\}
\end{align}
Analogous expressions result for the pseudo scalar case by
observing $\bar u(P) \gamma^\mu \gamma_5 u(P) = 2 S^\mu$.

The situation in the non-forward case is somewhat more complicated
and, therefore, will be omitted here. There, not only the Dirac but also
the Pauli structure has to be taken into account whose general form
coincides with the expression (\ref{NLO2}) with ${\cal K}^\mu_D$
being replaced by ${\cal K}^\mu_P$ but, as can be proven by
using the hadrons equation of motion, in accordance with Eq.~(\ref{DDT}) 
both the Dirac and Pauli structures mix with each other.

Finally, let us point to the fact that expressions similar to 
Eq.~(\ref{SNLO2}) 
have been obtained in Ref.~\cite{BB91} for
inclusive particle production in $e^+e^-$--annihilation. This work is based
on a procedure \cite{BB88} which, in order to obtain higher twists, avoids 
the consideration of group representations and, instead, makes use of the 
quark (and gluon) equations of motion. As already mentioned 
\cite{GLR99} in the case of vector (and tensor) operators by this approach 
the tracelessness of the (local) operators and, consequently, also the
definiteness of their geometric twist may be missing.


\subsection{Nonlocal off-cone operators of definite twist: The 
technique of harmonic polynomials}
Obviously, the $x-$dependence of the expressions (\ref{NLO2}) -- 
(\ref{lcNLO2}) is completely determined by the twist structure of 
the operator from which the matrix element has to be built. Therefore, 
let us derive these results directly from the related twist-2 operator.
But, instead starting from the nonlocal expression (\ref{OP2}) we use 
its local version being given by harmonic polynomials of order $n$ 
in $x$. This {\em polynomial technique}
uses the vector $x\in {\Bbb R}^4$ as a device for writing tensors with 
special symmetries in analytic form~\cite{BT77,Dobrev77,Dobrev82}. 
It has the advantage to be directly related to the irreducible tensor
representations of the Lorentz group. Its group theoretical background
as far as it is related to totally symmetric tensors has been given 
in Ref.~\cite{BT77}.

The local scalar operator according to (\ref{O^Gloc}) 
is given by 
\begin{align}
\label{o_n+1}
O_{n+1}(x) \equiv \bar{\psi}(0)(x\gamma)(x\Tensor D)^n \psi(0)
=\int \d^4 q \,\big(\bar{\psi}\gamma^\mu\psi\big)(q)\,x_\mu\,(xq)^n,
\end{align}
which shows that, in this connection, $q$ formally replaces the
covariant derivative $\Tensor D$ sandwiched between the quark operators.  
This local operator has a (finite) twist decomposition whose complete
series will be given later on, cf.,~Eq.~(\ref{O_loc_i}).
Its twist-2 part is simply given by the harmonic 
polynomials~\cite{VK,BT77},
\begin{align}
\label{O_tw2_sca}
O^{\rm tw2}_{n+1}(x)=
\sum_{k=0}^{[\frac{n+1}{2}]}\frac{(-1)^k (n+1-k)!}{4^k k!(n+1)!}\, x^{2k}
\square^k O_{n+1}(x)
\equiv {\cal P}^{(2)}_{n+1}(x^2,\pd^2)\,O_{n+1}(x),
\end{align}
being characterized by traceless, totally symmetric tensors of rank $n+1$
whose indices are completely contracted by $x^{\mu_1} \cdots x^{\mu_{n+1}}$.
They obey the condition of harmonicity: $\square O^{\rm tw2}_{n+1}(x) =0$.
 
Performing the derivatives,
\begin{align}
x^{2k}\square^k (\gamma x) (qx)^n=
\frac{n!}{(n-2k)!}\, (\gamma x) (q^2 x^2)^k (qx)^{n-2k}\theta_{n-2k}
+\frac{2k\,n!}{(n+1-2k)!}\,  x^2 (\gamma q) (q^2 x^2)^{k-1} (qx)^{n+1-2k}
\theta_{n+1-2k},
\end{align}
and using the series expansion of the Gegenbauer polynomials
(see, e.g.,~Ref.~\cite{PBM}, Appendix II.11),
\begin{align}
\label{GB10}
C_n^\nu(z)=\frac{1}{(\nu-1)!}\sum_{k=0}^{[\frac{n}{2}]}
\frac{(-1)^k(n-k+\nu-1)!}{k!(n-2 k)!}\,(2z)^{n-2k},
\end{align}
we obtain 
\begin{align}
\label{O2_n+1}
O^{\text{tw 2}}_{n+1}(x)
&=
\frac{1}{n+1}
\int\!\d^4 q\, \big(\bar{\psi}\gamma^\mu\psi\big)(q)
\left\{ x_\mu
\left(\hbox{\large$\frac{1}{2}$}\sqrt{q^2 x^2}\right)^{\!n}
\!C_n^2\bigg(\frac{qx}{\sqrt{q^2 x^2}}\bigg)
-\hbox{\large$\frac{1}{2}$} q_\mu\,x^2\,  
\left(\hbox{\large$\frac{1}{2}$}\sqrt{q^2 x^2}\right)^{\!n-1}
\!C_{n-1}^2\bigg(\frac{qx}{\sqrt{q^2 x^2}}\!
\bigg)
\right\}.
\end{align}
Again, some remarks are in order: 
\begin{itemize}
\item{
The factor $1/(n+1)$ in front of the integral results from the 
equality $x_\mu (xq)^n = (1/n+1) \pd_\mu^q (xq)^{n+1}$; it
corresponds to the normalization of the decomposition of
the local operators into irreducible representations, i.e., to the
normalization of the associated Young operators. In the nonlocal
formulation it leads to the $t-$integration in Eq.~(\ref{nfme}),
cf.,~also,~Ref.~\cite{GLR99}.}
\item{
The (discrete) $\theta-$functions
guarantee that both the $k-$summations do not contain undefined factorials
thus correctly leading to the corresponding Gegenbauer polynomials.
In the following this is circumvented by the convention that Gegenbauer
polynomials of negative order are equal to zero.}
\item{
Obviously, Eq.~(\ref{O2_n+1}) is the analytic continuation from
Euclidean spacetime --- where the Gegenbauer polynomials obey the
well-known orthonormality relations within the region $-1 \leq z \leq 1$
 --- to the Minkowski spacetime where
the arguments of the square roots and of the Gegenbauer polynomials,
depending on whether $q^2$ and/or $x^2$ are space-like or time-like,
may be imaginary and, furthermore, could take values outside 
that region. This, however, has no influence on the validity of
the above result since these polynomials being entire analytic functions
are well defined on the whole complex plane.}
\item{
Notice that the second term
in Eq.~(\ref{O2_n+1}) is due to the exterior $x-$dependence of the scalar
operator. Being proportional to $x^2$ it vanishes on the light-cone. 
Furthermore, on the light-cone only the highest power $z^n$
of the Gegenbauer polynomial $C_n^2(z)$ survives and, as it should
be because of (\ref{O_tw2_sca}), leads exactly to the expression 
(\ref{o_n+1}) with $x\rightarrow \lcx$!}
\item{
Finally, according to the above mentioned interpretation of $q$ as some kind
of `operator symbol' within the Fourier integral, Eq.~(\ref{O2_n+1}) 
operationally has to be understood as if the polynomial inside the 
curly brackets had been written in terms of $\Tensor D$ instead of $q$ and 
inserted into the operator $\bar\psi(0) \gamma^\mu \psi(0)$.}
\end{itemize}

Now, let us remind the generating function of the Gegenbauer polynomials
(see, e.g.,~Ref.~\cite{PBM}, Eq.~II.5.13.1.3),
\begin{align}
\sum_{n=0}^\infty\frac{a^n}{(2\nu)_n}\, C^\nu_n(z)=
\Gamma\left(\nu+\hbox{\large$\frac{1}{2}$}\right) 
\left(\frac{a}{2}\sqrt{1-z^2}\right)^{1/2-\nu}
J_{\nu-1/2}\left(a\sqrt{1-z^2}\right) \e^{za},
\end{align}
where 
$(2\nu)_n = 2\nu (2\nu +1) \ldots (2\nu+n-1) =
\Gamma(n+2\nu)/\Gamma(2\nu)$ is the Pochhammer symbol. 
Choosing $z=(qx)/\sqrt{q^2 x^2}$ and $a=\ii\sqrt{q^2 x^2}/2$, the local 
scalar operators of twist--2, Eq.~(\ref{O2_n+1}), can be summed up 
according to Eq.~(\ref{O^Gint}) to the bilocal scalar operator of twist--2:
\begin{align}
\label{XYZ}
O^{\rm tw2}(\kappa x,-\kappa x)=
\sum_{n=0}^\infty \frac{(\ii\kappa)^n}{n!}\, O^{\rm tw2}_{n+1}(x)
&=\sqrt{\pi}\int\!\d^4 q\, \big(\bar{\psi}\gamma^\mu\psi\big)(q)
\left\{ x_\mu\left(2+x\pd\right)
-\hbox{\large$\frac{1}{2}$}\,\ii\kappa q_\mu x^2 \right\}
\left(3+x\pd\right)
\nonumber\\&\qquad\qquad\times
\left(\kappa\sqrt{(qx)^2-q^2 x^2}\right)^{-3/2}
\!J_{3/2}\left(\hbox{\large$\frac{\kappa}{2}$}\sqrt{(qx)^2-q^2 x^2}\right)
\e^{\ii\kappa qx/2}.
\end{align}
Here, the homogeneous derivations $(c+x\pd)$ are required to compensate 
for some extra factors $(c+n)$ which are necessary in order to by able to
introduce the Pochhammer symbols in the denominator. Obviously, for the 
second term in Eq.~(\ref{O2_n+1}), after shifting $n \rightarrow n+1$ 
in the series over $n$, only one additional factor is required. ---
Also here $q$ has to be considered as a symbol replacing
the covariant derivatives 
sandwiched between the quark operators.

Now, let us discuss the matrix elements of the twist-2 operators
(\ref{O2_n+1}) and (\ref{XYZ}) starting with the forward case $P_1=P_2=P$.
The $q-$integral representation of the local twist-2 operator, observing
the symmetry in $q$ and $x$ of the twist-2 projection of 
 $(qx)^{n+1}$,
\begin{align}
{\cal P}^{(2)}_{n+1}(x^2,\pd^2)\, (xq)^{n+1}
=
 \left(\hbox{\large$\frac{1}{2}$}\sqrt{q^2 x^2}\right)^{\!n+1}
 C_{n+1}^1\bigg(\frac{qx}{\sqrt{q^2 x^2}}\bigg),
 \nonumber
 \end{align}
may be rewritten as follows
\begin{align}
O^{\text{tw 2}}_{n+1}(x) 
&=\frac{1}{n+1}
\int\!\d^4 q\, \big(\bar{\psi}\gamma^\mu\psi\big)(q)\,
\frac{\pd}{\pd q^\mu} \bigg\{
\left(\hbox{\large$\frac{1}{2}$}\sqrt{q^2 x^2}\right)^{\!n+1}
 C_{n+1}^1\bigg(\frac{qx}{\sqrt{q^2 x^2}}\bigg)\bigg\}
\nonumber\\
&=
\Big\{{\cal P}^{(2)}_{n+1}(x^2,\pd^2) 
\big(x^{\mu_1} \ldots x^{\mu_{n+1}}\big)\Big\}
\frac{1}{n+1}
\int\!\d^4 q\, \big(\bar{\psi}\gamma^\mu\psi\big)(q)\,
\frac{\pd}{\pd q^\mu} 
\big(q^{\mu_1} \ldots q^{\mu_{n+1}}\big)
\nonumber\\
&=
\Big\{{\cal P}^{(2)}_{n+1}(x^2,\pd^2) 
\big(x^{\mu_1} \ldots x^{\mu_{n+1}}\big)\Big\}
\frac{1}{n+1}
\int\!\d^4 q\, \big(\bar{\psi}\gamma^\mu\psi\big)(q)\,
\frac{\pd}{\pd q^\mu} 
\Big\{{\cal P}^{(2)}_{n+1}(q^2,\pd_q^2) 
\big(q^{\mu_1} \ldots q^{\mu_{n+1}}\big)\Big\}.
\nonumber
\end{align}
Obviously, the integrand is an irreducible tensor operator 
of order $n+1$. Its forward matrix elements are given by the
reduced matrix element $f^{(2)}_{n}$ times the irreducible tensor
written in terms of the momentum $P_+=2P$. As the result we finally obtain
\begin{align}
\label{lo_2}
\langle P,S|O^{\text {tw 2}}_{n+1}(x)|P,S\rangle
&=
\frac{f^{(2)}_{n}}{n+1}
\big(\bar u(P)\gamma^\mu u(P)\big)\,
\frac{\pd}{\pd (2P^\mu)} 
\bigg\{\!\!
\left(\sqrt{P^2 x^2}\right)^{\!n+1}
C_{n+1}^1\bigg(\frac{Px}{\sqrt{P^2 x^2}}\bigg)\!
\bigg\}
\nonumber\\
&=
\frac{f^{(2)}_{n}}{n+1}
(2Px)\,\bigg\{\!\!
\left(\sqrt{P^2 x^2}\right)^{\!n}
\!C_n^2\bigg(\frac{Px}{\sqrt{P^2 x^2}}\bigg)
-\frac{P^2\,x^2}{(xP)}\,  
\left(\sqrt{P^2 x^2}\right)^{\!n-1}
\!C_{n-1}^2\bigg(\frac{Px}{\sqrt{P^2 x^2}}\!
\bigg)\!
\bigg\}.
\end{align}
Now, as usual, the reduced matrix elements $f^{(2)}_n$ are considered
as the moments of a distribution $f^{(2)}(z)$,
\begin{align}
f^{(2)}_n = \int_{-1}^{1} \d z z^n f^{(2)}(z).
\end{align}
Taking this into account we obtain, after summing up over $n$, the
following expression of the forward matrix elements of the non-local
twist-2 operators 
\begin{align}
\label{xyz}
\langle P,S|O^{\rm tw2}(\kappa x,-\kappa x)|P,S\rangle
&=
\bar u(P)(x\gamma) u(P)
\int^1_{-1}\d z\,f^{(2)}(z)
\left\{\left(2+x\pd\right)
-\ii\kappa z \frac{P^2 x^2}{(xP)} \right\}
\left(3+x\pd\right)
\nonumber\\&\qquad\qquad\times
\sqrt{\pi}\left(2\kappa z\sqrt{(Px)^2-P^2 x^2}\right)^{-3/2}
\!J_{3/2}\left({\kappa} z\sqrt{(Px)^2-P^2 x^2}\right)
\e^{\ii\kappa z Px},
\end{align}
which coincides with the expression (\ref{SNLO2}).

More generally, when non-forward matrix elements are taken the argumentation
is almost the same. But now, in the case of the local operators, there occur
$n+1$ different reduced matrix elements, $f^{(2)}_{n,m}$, related to the 
monomials $P_1^m \, P_2^{n-m},~0 \leq m \leq n$ which are obtained from 
$(P_2+P_1)^n$, and the corresponding non-forward matrix element reads: 
\begin{align}
\label{nflo_2}
\langle P_2,S_2|O^{\text{tw 2}}_{n+1}(x)|P_1,S_1\rangle
&=
\sum_{m=0}^n \binom{n}{m}
f^{(2)}_{n,m}
\big({\bar u}(P_2)\gamma_\mu u(P_1)\big)\,
{\cal P}^{(2)}_{n+1}(x^2,\pd^2) 
\Big\{x^\mu (xP_1)^m (xP_2)^{n-m}\Big\}.
\end{align}
Now let us rewrite the reduced matrix elements $f^{(2)}_{n,m}$ as 
the double moments of some double distribution $f^{(2)}(z_1,z_2)$,
\begin{align}
f^{(2)}_{n,m} = \int_{-1}^1 \d z_1 \int_{-1}^1 \d z_2\, 
z_1^m z_2^{n-m} f^{(2)}(z_1,z_2).
\end{align}
After resumming w.r.t. $n$ and rewriting $f^{(2)}(z_1,z_2) =
F^{(2)}_D(z_+,z_-)$ we finally arrive at the expression (\ref{NLO2}) for the
non-forward matrix element of the nonlocal operator 
$O^{\text {tw 2}}(\kappa x, -\kappa x)$.

Obviously, this second approach is more appropriate since, as an
intermediate step, also knowledge about the local operators of
definite twist is obtained which is related to the corresponding
moments of the double distributions. From this approach it becomes also
obvious that the mass corrections of different physical processes
are related to each other 
if they can be traced back to the same operator content.

In the following we use this polynomial technique to determine the
local as well as nonlocal quark operators of definite twist. Thereby,
we generalize Nachtmann's approach not only to non-forward matrix
elements of (non)local operators.
In addition we consider also the case of more general
tensor operators having nontrivial symmetry types.

\section{Totally symmetric nonlocal harmonic operators of any 
geometric twist}
\setcounter{equation}{0}

The determination of the {\em complete} twist decomposition of non-local 
operators into the infinite tower of harmonic operators of definite twist 
is possible only for the case when their $n-$th moments are related to 
irreducible tensor representations of the Lorentz group having 
symmetry type $[n]$, i.e.,~being completely symmetric and traceless.
The corresponding group theoretical background underlying this polynomial 
technique has been formulated by Bargmann and Todorov~\cite{BT77}. 
There, also the projection property (\ref{proj}) is proven by explicit 
construction. In order to demonstrate the efficiency of the polynomial 
technique and because of their physical relevance we apply it to the complete 
twist decomposition of the scalar operators $N(x,-x),\, M(x,-x)$ and 
$O(x,-x)$, Eqs.~(\ref{N_nl}), (\ref{sM_nl}) and (\ref{sO_nl}), having 
minimal twist $\tau = 3$ and $\tau = 2$, respectively, as well as to 
the related vector and tensor operators. At the same time the general
aspects of the twist projection, Eqs.~(\ref{TWIST}) -- (\ref{proj}),
together with the related comments, will be exemplified.

\subsection{Scalar harmonic operators without external operations}

To begin with the simplest case let us study the local operators $N_n(x)$ 
of degree $n$ which are generating polynomials of symmetric tensors of 
degree $n$. The general formula for the decomposition of the 
local scalar operator $N_{n}(x)$ into its harmonic polynomials of definite
 twist $\tau=3+2j,\, j= 0, 1, \ldots ,$ reads~\cite{BT77}
\begin{align}
\label{N_tw_sc}
N_{n}(x)=
\sum_{j=0}^{[\frac{n}{2}]}
\frac{(-1)^j (n+1-2j)!}{4^j j!(n+1-j)!}\, x^{2j}\, 
N^{{\rm tw}(3+2j)}_{n-2j}(x),
\end{align}
where the harmonic operators of definite twist are defined by 
(see also Ref.~\cite{VK}, Eq.~9.3.2(3))
\begin{align}
\label{Ntg}
N^{{\rm tw}(3+2j)}_{n-2j}(x)=
\sum_{k=0}^{[\frac{n-2j}{2}]}
\frac{(-1)^k (n-2j-k)!}{4^k k!(n-2j)!}\, x^{2k}\square^k
\left(\square^{j} N_{n}(x)\right), 
\end{align}
they satisfy
\begin{align}
\square N^{{\rm tw}(3+2j)}_{n-2j}(x)=0.
\nonumber
\end{align}
Therefore, the polynomials of Eq.~(\ref{Ntg}) span
the space of homogeneous harmonic polynomials of degree $n-2j$.

After substituting $N_n(x) = \int \d q\, N(q) (xq)^n$
 into (\ref{Ntg}), and observing 
\begin{align}
\square^{k+j} \,(xq)^n = 
\frac{n!}{(n-2k-2j)!} (q^2)^{k+j} (xq)^{n-2k-2j},
\nonumber
\end{align}
we can rewrite
these harmonic operators in terms of Gegenbauer polynomials as follows:
\begin{align}
\label{Nttw_GP}
&N^{\text{tw}(3+2j)}_{n-2j}(x) =
\frac{n!}{(n-2j)!}
\int\!\d q\, N(q)\, 
q^{2j}\left(\hbox{\large$\frac{1}{2}$}\sqrt{q^2 x^2}\right)^{n-2j}
\!C_{n-2j}^1\left(\frac{qx}{\sqrt{q^2 x^2}}\right).
\end{align}
Here, the factorials in front of the integral are normalization
coefficients and the $q-$integration introduces a superposition
of the Fourier transforms $N(q)$ times $q^{2j}$ with coefficients 
being (two-sided) harmonic polynomials, 
\begin{align}
h^1_{n-2j}(q|x) = \big(\hbox{\Large$\frac{1}{2}$}
\sqrt{q^2 x^2}\big)^{n-2j}
\!C_{n-2j}^1\big({qx}/{\sqrt{q^2 x^2}}\big)
\qquad
{\rm with}
\qquad
\square_x\; h^1_{n-2j}(q|x) = 0 = \square_q\; h^1_{n-2j}(q|x).
\end{align}

The resummation of the moments $N_n(x)$ according to 
$N(x,-x) = \sum_{n=0}^\infty ({\ii^n}/{n!}) N_n(x)$ is obtained by using the 
well-known integral representation of Euler's beta function,
\begin{align}
\label{beta}
B(n,m)=\frac{\Gamma(n)\Gamma(m)}{\Gamma(n+m)}=
\int_0^1\d t\, t^{n-1} (1-t)^{m-1}.
\end{align}
The (infinite) twist decomposition of the nonlocal scalar operator 
$N(x,-x)$  into nonlocal harmonic operators $N^{{\rm tw}(3+2j)}(x,-x)$
of twist $\tau=3+2j,\, j= 0,1, 2, \ldots,$ reads
(remind that the summation over $n$ for 
$N_{n-2j}^{{{\rm tw}(3+2j)}}$ starts at $n = 2j$), 
\begin{align}
\label{N_twn_sc}
N(x,-x)= N^{{\rm tw}3}(x,-x)
+ \sum_{j=1}^{\infty}
\frac{(-1)^j x^{2j}}{4^j j!(j-1)!}
\int_0^1\d t\, t\,(1-t)^{j-1}\, 
N^{{\rm tw}(3+2j)}(tx,-tx),
\end{align}
with 
\begin{align}
\label{N^(3+2j)}
N^{{\rm tw}(3+2j)}(x,-x)
&=\sqrt{\pi}\int\!\d q\, N(q)\, 
(- q^2)^{j}\left(1+q\pd_q\right)
\left(\sqrt{(qx)^2-q^2 x^2}\right)^{-1/2}
\!J_{1/2}\left(\hbox{\large$\frac{1}{2}$}\sqrt{(qx)^2-q^2 x^2}\right)
\e^{\ii  qx/2}.
\end{align}
Let us remark that, for notational simplicity, in Eq.~(\ref{N_twn_sc})
we wrote the nonlocal harmonic operators for $j \geq 1$ 
without including the integration over $t$  
which is due to the normalization of the local harmonic operators,
cf.~Eqs.~(\ref{N_tw_sc}) and (\ref{Nttw_GP}). The normalization
of the nonlocal harmonic operators may be read off from the terms
in front of the $t-$integral in Eq.~(\ref{N_twn_sc}).
Let us point also to the remarkable fact that the property of
harmonicity is independent of $j$; this obtains immediately from the 
fact that the following expressions are (two-sided) harmonic functions,
\begin{align}
\label{harm}
\hspace{-.5cm}
{\cal H}_1(q|x) = \sqrt{\pi}\left(\sqrt{(qx)^2-q^2 x^2}\right)^{-1/2}
\!J_{1/2}\left(\hbox{\large$\frac{1}{2}$}\sqrt{(qx)^2-q^2 x^2}\right)
\e^{\ii  qx/2}
\qquad
{\rm with}
\qquad
\square_x\, {\cal H}_1(q|x) = 0 = \square_q\,{\cal H}_1(q|x).
\end{align}
Furthermore, we observe that the factor $(1+q\pd/\pd q)$ in 
Eq.~(\ref{N^(3+2j)}) may be changed into $(1+x\pd/\pd x)$ after which it
can be taken outside the $q-$integration and, in Eq.~(\ref{N_twn_sc}),
could be changed into $(1+t\pd/\pd t)$; if desired, this could
be used for a partial integration.

The resulting expression (\ref{N^(3+2j)}) can be rewritten in
a form which, in the case $j=0$, has been already introduced 
by \cite{BB91,Ball99}. Namely, using
\begin{align}
z^{-n-1/2} J_{n+1/2}(z) = (-1)^n \sqrt{\frac{2}{\pi}} 
\Big(\frac{1}{z} \frac{\d}{\d z}\Big)^n
\Big(\frac{\sin z}{z}\Big),
\end{align}
we obtain
\begin{align}
N^{{\rm tw}(3+2j)}(x,-x) 
=
\int \d q\, N(q)\, (-q^2)^j\,
&\frac{2}{\sqrt{(qx)^2-q^2 x^2}}\, q\pd_q \,
\Big( \sin \big(\hbox{\large$\frac{1}{2}$}
\sqrt{(qx)^2-q^2 x^2}\big) \, \e^{\ii qx/2} \Big)
\nonumber\\
=
\int \d q\, N(q)\, (-q^2)^j\,
&\frac{1}{2\sqrt{(qx)^2-q^2 x^2}}
\bigg( 
\big(qx + \sqrt{(qx)^2-q^2 x^2}\big)\,
\e^{(\ii/2)\big(qx + \sqrt{(qx)^2-q^2 x^2}\big)}
\nonumber\\
&\qquad\qquad\qquad\quad
-
\big(qx - \sqrt{(qx)^2-q^2 x^2}\big)\,
\e^{(\ii/2)\big(qx - \sqrt{(qx)^2-q^2 x^2}\big)}
\bigg).
\end{align}
Analogous results may be obtained also in the more
complicated cases which will be considered below.

\subsection{Scalar harmonic operators with external operations}

Now, let us consider the case of $O(x,-x) = x^\alpha O_\alpha(x,-x)$, 
where an additional power of $x$ occurs through contraction of the vector
operator with $x^\alpha$. Therefore, the formulae (\ref{N_tw_sc}) and 
(\ref{Ntg}) are to be written down for the scalar operator 
$O_{n+1}(x)= \int \d q\, x_\mu O^\mu(q) (xq)^n$, 
i.e.,~by replacing $n$ by $n+1$ within these expressions for the 
decomposition into harmonic operators of any twist $\tau=2+2j$.
Then the decomposition of the scalar local operators 
$O_{n+1}(x)$ reads:
\begin{align}
\label{O_loc_i}
O_{n+1}(x)=\sum_{j=0}^{[\frac{n+1}{2}]}
\frac{(-1)^j (n+2-2j)!}{4^j j!(n+2-j)!}\, x^{2j}\, 
O^{{\rm tw}(2+2j)}_{n+1-2j}(x),
\end{align}
where the local harmonic operators of twist $\tau = 2 + 2j$ are given by 
\begin{align}
\label{Otg}
O^{{\rm tw}(2+2j)}_{n+1-2j}(x)=
\sum_{k=0}^{[\frac{n+1-2j}{2}]}\frac{(-1)^k (n+1-2j-k)!}{4^k k!(n+1-2j)!}\, 
x^{2k}\square^k\left(\square^{j} O_{n+1}(x)\right). 
\end{align}
After performing the differentiations $\square^{k+j} O_{n+1}(x)$ 
we can rewrite
these harmonic operators in terms of Gegenbauer polynomials as follows:
\begin{align}
\label{Ottw_GP}
\hspace{-.3cm}
O^{\text{tw}(2+2j)}_{n+1-2j}(x)& =
\frac{n!}{(n+1-2j)!}
\int\!\d q\, O^\mu(q)\, q^{2j}\,\bigg\{ x_\mu 
\left(\hbox{\large$\frac{1}{2}$}\sqrt{q^2 x^2}\right)^{n-2j}
\!C_{n-2j}^2\bigg(\frac{qx}{\sqrt{q^2 x^2}}\bigg)\\
\hspace{-.3cm}
&\qquad
- \frac{1}{2} q_\mu  x^2
\left(\hbox{\large$\frac{1}{2}$}\sqrt{q^2 x^2}\right)^{n-1-2j}
\!C_{n-1-2j}^2\bigg(\frac{qx}{\sqrt{q^2 x^2}}\bigg)
+2j \, \frac{q_\mu}{q^2}
\left(\hbox{\large$\frac{1}{2}$}\sqrt{q^2 x^2}\right)^{n+1-2j}
\!C_{n+1-2j}^1\bigg(\frac{qx}{\sqrt{q^2 x^2}}\bigg)
\bigg\}, \nonumber
\end{align}
where the additional terms proportional to $q_\mu$ occur 
because of the appearance of derivatives with respect to $(x\gamma)$.
The change in the order of the Gegenbauer polynomials and, consequently,
in the order of the Bessel functions and the accompanying factors
$\sqrt{(qx)^2-q^2 x^2}$ occurring below, is also due to that fact. 
Of course, for $j=0$ we re-obtain the expression (\ref{O2_n+1})

Resumming with respect to $n\ (\,\geq 2j \pm 1$ or $2j,$ respectively) 
and, again, using the representation (\ref{beta}) of Euler's beta function
we obtain the following (infinite) twist decomposition
\begin{align}
\label{O_twn_sc}
O(x,-x)=O^{{\rm tw}2}(x,-x)
+\sum_{j=1}^{\infty}
\frac{(-1)^j x^{2j}}{4^j j!(j-1)!}
\int_0^1\d t\, t\, (1-t)^{j-1}\, 
O^{{\rm tw}(2+2j)}(tx,-tx),
\end{align}
with 
\begin{align}
\label{O^(2+2j)}
O^{{\rm tw}(2+2j)}(x,-x)
&=\sqrt{\pi}\int\!\d q\, O^\mu(q)\,(- q^2)^{j}
\bigg\{
\left( x_\mu(2+q\pd_q)
-\hbox{\large$\frac{1}{2}$}\, \ii q_\mu x^2  
\right)\nonumber\\
&\qquad\qquad\times
\left(3+q\pd_q\right)\left(\sqrt{(qx)^2-q^2 x^2}\right)^{-3/2}
\!J_{3/2}\left(\hbox{\large$\frac{1}{2}$}\sqrt{(qx)^2-q^2 x^2}\right)
\nonumber\\
&\quad\; - 2j\, \frac{ \ii q_\mu}{ q^{2}} \left(1+q\pd_q\right)
\left(\sqrt{(qx)^2-q^2 x^2}\right)^{-1/2}
\!J_{1/2}\left(\hbox{\large$\frac{1}{2}$}\sqrt{(qx)^2-q^2 x^2}\right)
\bigg\}\,\e^{\ii qx/2}.
\end{align}
Again, by convention, the resummed nonlocal harmonic operators 
$O^{{\rm tw}(2+2j)}(tx,-tx)$ of twist $\tau=2+2j, \, j = 0,1,2,\cdots,$ 
are obtained by changing any $x \rightarrow tx$. 
Also here analogous comments are in order as for the simpler case of the 
operator $N(x,-x)$. Harmonicity of the nonlocal operators (\ref{O^(2+2j)}), 
$\square O^{{\rm tw}(2+2j)}(x,-x) = 0$, holds by construction and, again,
it is independent of $j$; namely, it holds separately for the last, 
$j$--dependent term due to the equality (\ref{harm}) and, therefore, 
must be fulfilled also for the combination of the first and second term.
Obviously, for $j=0$ we recover the expression (\ref{XYZ}) .

\medskip

Secondly, let us consider the scalar operator
$M(x,-x) =  x^\nu \pd^\mu M_{[\mu\nu]} (x,-x)$ resulting from the 
skew tensor operator $M_{[\mu\nu]} (x,-x)$. This operator is
governed by totally symmetric tensors since the skew tensors  
$M_{[\mu\nu]n}$ having symmetry type $[n+1,1]$ are composed with another 
skew tensor $x^{[\nu}\pd^{\mu]}$ having symmetry type $[1,1]$ and the 
Clebsch--Gordan series of their direct product consists only of one
symmetry type, $[n]$.

The local operators are decomposed according to formulae
(\ref{N_tw_sc}) and (\ref{Ntg}). However, because of the
special `external' structure the resulting expressions
simplify considerably. Namely, after performing the differentiations 
\begin{align*}
 \square^{k+j} \pd^\mu\sigma_{\mu\nu} x^\nu (xq)^n
= \big({n!}/{(n-1-2j)!}\big) \big(q^\mu\sigma_{\mu\nu} x^\nu\big)
 q^{2j} (q^2)^k (qx)^{n-1-2k-2j},
\end{align*}
 we arrive at the following harmonic 
operators in terms of Gegenbauer polynomials:
\begin{align}
\label{Mttw_GP}
&M^{\text{tw}(3+2j)}_{n-2j}(x) =
\frac{n!}{(n-1-2j)!}
\int\!\d q\, M_{[\mu\nu]}(q)\, q^\mu x^\nu\,q^{2j}\,
\left(\hbox{\large$\frac{1}{2}$}\sqrt{q^2 x^2}\right)^{n-1-2j}
\!C_{n-1-2j}^2\left(\frac{qx}{\sqrt{q^2 x^2}}\right).
\end{align}

Now, resumming with respect to $n\, (\,\geq 2j+1)$ we finally get the 
infinite twist decomposition:
\begin{align}
\label{M_twn_sc}
M(x,-x)=M^{{\rm tw}3}(x,-x)
+\sum_{j=1}^{\infty}
\frac{(-1)^j x^{2j}}{4^j j!(j-1)!}
\int_0^1\d t\, t\, (1-t)^{j-1}\, 
M^{{\rm tw}(3+2j)}(tx,-tx),
\end{align}
with 
\begin{align}
\label{M^(2+2j)}
M^{{\rm tw}(3+2j)}(x,-x)
&=\sqrt{\pi}\int\!\d q\, 
M_{[\mu\nu]}(q)\,(- q^2)^{j}\,\ii q^\mu x^\nu
\nonumber\\
&\qquad\qquad\times
\left(2+q\pd_q\right)\left(3+q\pd_q\right)
\left(\sqrt{(qx)^2-q^2 x^2}\right)^{-3/2}
\!J_{3/2}\left(\hbox{\large$\frac{1}{2}$}\sqrt{(qx)^2-q^2 x^2}\right)
\e^{\ii qx/2}.
\end{align}
Again, harmonicity of the  nonlocal operators 
$M^{{\rm tw}(3+2j)}(x,-x)$ of twist $\tau=3+2j, \, j = 0,1,2,\cdots,$ 
is fulfilled without taking care of $j$.


\subsection{Harmonic vector and tensor operators
related to totally symmetric local operators}


The prescribed procedure may be used also for the twist decomposition 
of any totally symmetric (tensor) operator 
with an arbitrary number of free tensor indices which, by construction, are
obtained from the corresponding scalar operators simply by applying
appropriate derivatives with respect to $x$. 

The simplest of these operators, whose tower of infinite twist part starts 
with $\tau =2$ and which contains the leading contributions to virtual 
Compton scattering is the operator $O^{\sf S}_\alpha (x,-x)$. Its symmetry 
type $\sf S$ is characterized by the Young frames
$[n+1],\ 1\leq n\leq\infty$ (cf.~also Ref.~\cite{GLR99} where we denoted
that symmetry type by (i)).
Its moments, $O^{\sf S}_{\alpha n}(x)$,  are obtained from the local 
scalar operators (\ref{O_loc_i}) of degree $n+1$ by applying 
$(1/(n+1)) \pd_\alpha$.  For the moment leaving aside the factor 
$(n+1)^{-1}$ we get
\begin{align}
\label{O_tw_sc}
\pd_\alpha O_{n+1}(x)
=\sum_{j-1=0}^{[\frac{n+1}{2}]}
\frac{(-1)^j (n+2-2j)!}{4^j (j-1)!(n+2-j)!}\, 2x_\alpha\, x^{2(j-1)}\, 
O^{{\rm tw}(2+2j)}_{n+1-2j}(x)
+
\sum_{j=0}^{[\frac{n+1}{2}]}
\frac{(-1)^j (n+2-2j)!}{4^j j!(n+2-j)!}\, x^{2j} 
\left(\pd_\alpha O^{{\rm tw}(2+2j)}_{n+1-2j}(x)\right).
\nonumber
\end{align}
Using the representation (\ref{beta}) of the beta function
both sums may be rewritten as
\begin{gather}
- \hbox{\large$\frac{1}{2}$}\,x_\alpha \sum_{j=0}^{[\frac{n+1}{2}]}
\frac{(-1)^j (n-2j)!}{4^j j!(n+1-j)!}\,  x^{2j}\, 
O^{{\rm tw}(4+2j)}_{n-1-2j}(x)
=
- \hbox{\large$\frac{1}{2}$}\,
x_\alpha \sum_{j=0}^{[\frac{n-1}{2}]}
\frac{(-1)^j\,x^{2j}}{4^j (j!)^2}  
\int_0^1\d t\, t\, (1-t)^j\,
O^{{\rm tw}(4+2j)}_{n-1-2j}(tx),
\nonumber\\
\sum_{j=0}^{[\frac{n+1}{2}]}
\frac{(-1)^j (n+2-2j)!}{4^j j!(n+2-j)!}\, x^{2j} 
\pd_\alpha O^{{\rm tw}(2+2j)}_{n+1-2j}(x)
=
\pd_\alpha O^{{\rm tw}\ 2}_{n+1}(x)
+
\sum_{j=1}^{[\frac{n+1}{2}]}
\frac{(-1)^j\,x^{2j}}{4^j j!(j-1)!}  
\int_0^1\d t\, t\, (1-t)^{j-1}
\left(\pd_\alpha O^{{\rm tw}(2+2j)}_{n+1-2j}(tx)\right),
\nonumber
\end{gather}
respectively, where
$O^{{\rm tw}(4+2j)}_{n-1-2j}(x)$ is given by Eq.~(\ref{Ottw_GP})
with $j \rightarrow j+1$. From this it becomes obvious that, 
beginning with twist-4, one obtains two different contributions
of the same twist,
namely a vector and a scalar part.

Now, putting together these terms and 
resumming over $n$, thereby representing the normalizing coefficient
$1/(n+1)$ as $\int_0^1 \d \lambda\, \lambda^n$, we obtain
\begin{align}
\label{O_v_sym}
O^{\sf S}_\alpha(x, -x)
&=
\pd_\alpha\int_0^1 \frac{\d\lambda}{\lambda}
O(\lambda x, - \lambda x)
\nonumber\\
&= 
\int_0^1\,\d\lambda\,
\bigg\{\big(\pd_\alpha\,O^{\mathrm tw2}\big)(\lambda x,-\lambda x)
+
\sum_{j=1}^\infty
\frac{(-1)^j x^{2j}\lambda^{2j}}{4^j j!(j-1)!}\,  
\int_0^1\d t\, t\, (1-t)^{j-1}\,
\big(\pd_\alpha\, O^{{\mathrm tw}(2+2j)}\big)(\lambda tx, -\lambda tx)
\nonumber\\
&\qquad\qquad\qquad\qquad\qquad\qquad
-\hbox{\large$\frac{1}{2}$}\,x_\alpha
\sum_{j=0}^\infty
\frac{(-1)^j x^{2j}\lambda^{2j+1}}{4^j (j!)^2}\,  
\int_0^1\d t\, t\, (1-t)^{j}\,
O^{{\mathrm tw}(4+2j)}(\lambda tx, -\lambda tx)
\bigg\},
\end{align} 
where 
$O^{{\mathrm tw}(4+2j)}(x, -x)\equiv O^{{\mathrm tw}(2+2(j+1))}(x, -x)$
and $O^{{\mathrm tw}(2+2j)}(x, -x)$ are given by Eq.~(\ref{O^(2+2j)}).
Both the integrals over $\lambda$ and $t$ are due to normalizations
and, therefore, it depends on the personal taste if they are to be
included into the definition of the harmonic operators of twist $\tau$
or if they are considered as part of the twist decomposition.
It is interesting to note that in Eq.~(\ref{O_v_sym}) only the twist-2 
vector and the twist-4 scalar operator survives on the light-cone.
All other higher twist operators for $j\geq1$ are cancelled due to the 
factor $x^2$.  

In the same manner we are able to define the twist decomposition of 
(completely symmetric) tensor operators which result from the expression
\begin{align}
O^{\sf S}_{(\alpha_1\alpha_2\cdots\alpha_r)n}(x)
=
\frac{\pd_{\alpha_1}}{(n+1)} \cdots
\frac{\pd_{\alpha_r}}{(n+r)}
\sum_{j=0}^{[\frac{n+r}{2}]}
\frac{(-1)^j (n+1+r-2j)!}{4^j j!(n+1+r-j)!}\, x^{2j}\, 
O^{{\rm tw}(2+2j)}_{n+r-2j}(x)\ .
\nonumber
\end{align}
The general procedure is obvious from the consideration of the vector
operator. Completely analogous to that case with only one derivative we 
would obtain a finite number of different operators of a given twist $\tau$.

However, in the case of operators having more complicated symmetry type,
e.g.,~being characterized by the symmetry type $[n+1,1]$, which are
necessary for the consideration, e.g.,~of skew symmetric tensor operators
like $M_{[\alpha\beta]}(x,-x)$, a theory which generalizes the formalism
of Ref.~\cite{BT77} is not available up to now. There exist only results
for the lowest values of $\tau$ which, however, are sufficient for the
consideration of the physically relevant cases. They will be considered
in the next chapter.

\section{Harmonic quark-antiquark operators of geometric twist 2 and 3}
\setcounter{equation}{0}

In this section we determine the harmonic quark-antiquark operators
up to twist $\tau = 2$ and $3$ which are relevant for the generalization
of the parton distributions of definite (geometric) twist \cite{GL01}
to the double distributions appearing in the parametrization in the
non-forward matrix elements of the corresponding non-local operators
off the light-cone.
Unfortunately, in the case of non-trivial Young frames where, up to now, 
no general group theoretical study being comparable with 
that of Ref.~\cite{BT77}. However, harmonic twist operators 
which are sufficient for the physically relevant off-cone operators 
have been obtained in Refs.~\cite{GLR99,GL00a}. They allow for the 
consideration of the vector and skew tensor operators up to twist 3.
According to the procedure of Sect.~II.C we start with their moments,
i.e.,~the local off-cone harmonic operators of definite twist. These
(pseudo) scalar, (axial) vector and skew tensor operators  
have been already determined (implicitly) in Ref.~\cite{GLR99}, Appendix B.
They are obtained by applying onto the harmonic tensor operators -- 
which ensure tracelessness -- the appropriate differential operators
which take care of the required symmetry types.

\subsection{Local harmonic operators: 
Series representation}

To begin with we introduce the local harmonic operators of geometric 
twist-2 and twist-3 for both the chiral-even (axial) vector operators,
\begin{align}
\label{O_ent}
O_{\alpha n}(x)&=\bar{\psi}(0)\gamma_{\alpha}(x\Tensor D)^n \psi(0),\\
\label{O5_ent}
O_{5\alpha n}(x)&
=\bar{\psi}(0)\gamma_{\alpha}\gamma_5 (x\Tensor D)^n \psi(0),
\end{align}
and the chiral-odd scalar and skew tensor operators,
\begin{align}
\label{N_ent}
N_{n}(x)&=\bar{\psi}(0)(x\Tensor D)^n \psi(0),\\
\label{M_ent}
M_{[\alpha\beta]n}(x)&
=\bar{\psi}(0)\sigma_{\alpha\beta}(x\Tensor D)^n \psi(0),
\end{align}
together with their related scalar and vector operators
\begin{align}
O_{n+1}(x)=x^\alpha O_{\alpha n}(x),
\qquad
M_{\mu n+1}(x)=x^\nu M_{[\mu\nu] n}(x),
\qquad
M_{n}(x)=\pd^\mu M_{\mu n+1}(x).
\end{align}

First we consider the vector operators $O_{\alpha n}(x)$ which 
allow for two different symmetry types \cite{GLR99,GL00a}. 
The irreducible vector operators $O^{\rm tw2}_{\alpha n}(x)$
of Lorentz type $(\frac{n+1}{2},\frac{n+1}{2})$ have symmetry types
$[n+1]$ and can be taken over from the previous section. 
On the other hand, the twist-3 vector operators transform according to  
$(\frac{n+1}{2},\frac{n-1}{2})\oplus(\frac{n-1}{2},\frac{n+1}{2})$
and have symmetry types $[n,1]$. They may be summed up to the lowest
term of another infinite tower of non-local operators of twist
$\tau = 3, 4, \ldots.$
The conditions of tracelessness for harmonic vector operators are as 
follows: 
\begin{align}
\label{traceless}
\square O^{\rm tw\,\tau}_{\alpha n}(x)=0,
\qquad 
\pd^\alpha O^{\rm tw\,\tau}_{\alpha n}(x)=0
\qquad
{\rm for}
\qquad
\tau=2,3.
\end{align}

For the sake of completeness and in order to introduce a new 
constructive element let us reconsider the twist-2 vector operator.
It is obtained from the scalar twist-2 operator, cf.~Eq.~(\ref{O_loc_i}),  
\begin{align}
O^{{\rm tw} 2}_{n+1}(x)=
\sum_{k=0}^{[\frac{n+1}{2}]}\frac{(-1)^k (n+1-k)!}{4^k k!(n+1)!}\, 
x^{2k}\,\square^k O_{n+1}(x). 
\nonumber
\end{align}
by means of a partial differentiation, $\pd_\alpha$, together with 
the normalization factor $1/(n+1)$ as follows:
\begin{align}
\label{O_tw2}
O^{\rm tw2}_{\alpha n}(x)
&=\frac{1}{n+1}\, \pd_\alpha O^{\rm tw2}_{n+1}(x)
\nonumber\\
&=\frac{1}{n+1}
\sum_{k=0}^{[\frac{n}{2}]}\frac{(-1)^k (n-k)!}{4^k k!(n+1)!}\, x^{2k}\,
{\cal D}_\alpha(k)\, \square^k O_{n+1}(x).
\end{align}
Here, we introduced the operation
\begin{align}
\label{D}
{\cal D}_\alpha(k) = (k+1+x\pd)\pd_\alpha
-\hbox{\large$\frac{1}{2}$}\,x_\alpha\square
\end{align}
which is a generalization of the inner derivative (for the latter, 
cf.~Ref.~\cite{GL01} and, more detailed, Refs.~\cite{BT77,Dobrev77})
off the light-cone and its extension to arbitrary values of $k\geq 0$.
(A similar derivative in momentum space already has been introduced
in Ref.~\cite{BE76}.)
As a side remark we mention that with the help of this generalized 
derivative also symmetric tensor operators of rank 2 and
higher can easily be constructed.

The twist-3 vector operator which satisfies the conditions 
(\ref{traceless}) is, with the convention 
$A_{[\mu\nu]}\equiv\frac{1}{2}\left(A_{\mu\nu}-A_{\nu\mu}\right)$, 
\begin{align}
\label{O_tw3}
O^{\rm tw3}_{\alpha n}(x)
&=\frac{2}{n+1}\, x^\beta
\bigg\{\delta^\mu_{[\alpha}\pd_{\beta]}-\frac{1}{n+1}\, 
x_{[\alpha}\pd_{\beta]}\pd^{\mu}\bigg\}
\sum_{k=0}^{[\frac{n}{2}]}\frac{(-1)^k (n-k)!}{4^k k!n!}\, 
x^{2k}\,\square^k O_{\mu n}(x),
\nonumber\\
&=\frac{2}{n+1}
\sum_{k=0}^{[\frac{n-1}{2}]}
\frac{(-1)^k (n-1-k)!}{4^k k!n!}\, x^{2k} x^\beta
\Big\{\delta^\mu_{[\alpha}
{\cal D}_{\beta]}(k)
-\frac{1}{n+1}\, x_{[\alpha}\pd_{\beta]}
{\cal D}^\mu(k)
\Big\}\,\square^k
O_{\mu n}(x).
\end{align}

The twist-4 vector operators having symmetry type $[n+1]$ can be read off
from the general result of Sec.~III.C. However, there occur also 
twist-4 vector operators having symmetry type $[n,1]$ which
have been determined on the light-cone only \cite{GLR99,GL00a}.

The scalar twist-3 operator is of Lorentz type 
$(\frac{n}{2},\frac{n}{2})$ and also has been considered in
Sec.~III; it reads
\begin{align}
\label{N_tw3}
N^{\rm tw3}_{n}(x)=
\sum_{k=0}^{[\frac{n}{2}]}\frac{(-1)^k (n-k)!}{4^k k!n!}\, x^{2k}
\square^k N_{n}(x).
\end{align}

Now, let us give the decomposition of the skew tensor operator 
$M_{[\alpha\beta] n}(x)$ into its twist-2 and twist-3 parts.
This determination is simplified by the observation that
these operators are related to the vector operator 
$M_{\mu n+1}(x)=x^\nu M_{[\mu\nu] n}(x)$
and the scalar twist-3 operator $M_{n}(x)=\pd^\mu M_{\mu n+1}(x)$,
respectively.
The twist-2 tensor operator $M^{\rm tw2}_{[\alpha\beta] n}(x)$
transforms according to 
$(\frac{n+2}{2},\frac{n}{2})\oplus(\frac{n}{2},\frac{n+2}{2})$, whereas
the twist-3 operator which is obtained
from the trace terms of $M^{\rm tw2}_{[\alpha\beta] n}(x)$
is of Lorentz type $(\frac{n}{2},\frac{n}{2})$.
They have symmetry types $[n+1,1]$ and $[n]$, respectively,
and are given by
\begin{align}
\label{M_tw2}
M^{\rm tw2}_{[\alpha\beta] n}(x)
&=\frac{2}{n+2}
\bigg\{\delta^\mu_{[\alpha}\pd_{\beta]}-\frac{1}{n+2}\, 
x_{[\alpha}\pd_{\beta]}\pd^{\mu}\bigg\}
\sum_{k=0}^{[\frac{n+1}{2}]}\frac{(-1)^k (n+1-k)!}{4^k k!(n+1)!}\, 
x^{2k}\,\square^k M_{\mu n+1}(x)
\nonumber\\
&=\frac{2}{n+2}
\sum_{k=0}^{[\frac{n}{2}]}\frac{(-1)^k (n-k)!}{4^k k!(n+1)!}\, x^{2k}
\Big\{\delta^\mu_{[\alpha} {\cal D}_{\beta]}(k)
-\frac{1}{n+2}\, x_{[\alpha}\pd_{\beta]}{\cal D}^\mu(k)
\Big\}\,\square^k M_{\mu n+1}(x),
\\
\intertext{and}
\label{M_tw3a}
M^{\rm tw 3}_{[\alpha\beta] n}(x)&=\frac{2}{(n+2)}\,
x_{[\alpha}\widehat{M}^{\rm tw3}_{\beta] n-1}(x),
\\
\intertext{with}
\label{vM_tw3}
\widehat{M}^{\rm tw3}_{\beta n-1}(x)
&=\frac{1}{n}\, \pd_\beta 
\sum_{k=0}^{[\frac{n}{2}]}\frac{(-1)^k (n-k)!}{4^k k!n!}\, x^{2k}
\square^k M_{n}(x)
\nonumber\\
&=\frac{1}{n}
\sum_{k=0}^{[\frac{n-1}{2}]}\frac{(-1)^k (n-1-k)!}{4^k k!n!}\, x^{2k}\,
{\cal D}_\beta(k)\,\square^k M_{n}(x)
\end{align}

Besides the twist-3 operator (\ref{M_tw3a}) resulting from the trace
terms of the twist-2 operator (\ref{M_tw2}) there exists also a
genuine twist-3 operator having Lorentz type 
$\big(\frac{n}{2},\frac{n}{2}\big)$ and being governed
by the symmetry type $[n,1,1]$. It is given by the following expression
(with the convention $A_{\alpha\beta\gamma}\equiv \frac{1}{3} 
(A_{\alpha[\beta\gamma]}+A_{\beta[\gamma\alpha]}+A_{\gamma[\alpha\beta]})$):
\begin{align}
\label{M_tw3b}
\widetilde M^{\rm tw 3}_{[\alpha\beta] n}(x)
&=\frac{3}{n+2} x^\gamma
\bigg\{\delta^\mu_{[\alpha}\delta^\nu_{\beta}\pd_{\gamma]}
-\frac{2}{n}\, 
\delta^\mu_{[\alpha}x_\beta \pd_{\gamma]}\pd^{\nu}\bigg\}
\sum_{k=0}^{[\frac{n}{2}]}\frac{(-1)^k (n-k)!}{4^k k!n!}\, 
x^{2k}\,\square^k M_{[\mu\nu] n}(x)
\nonumber\\
&=\frac{3}{n+2}
\sum_{k=0}^{[\frac{n}{2}]}\frac{(-1)^k (n-1-k)!}{4^k k!n!}\,
x^{2k}\,x^\gamma
\Big\{\delta^\mu_{[\alpha}\delta^\nu_{\beta}{\cal D}_{\gamma]}(k)
-\frac{2}{n}\, 
\delta^\mu_{[\alpha}x_\beta \pd_{\gamma]}{\cal D}^{\nu}(k)\Big\}\, 
\square^k M_{[\mu\nu] n}(x).
\end{align}
By construction, Eqs.~(\ref{M_tw2}), (\ref{vM_tw3}) and (\ref{M_tw3b}) are 
(traceless) harmonic polynomials obeying
\begin{align}
\square M^{\rm tw\,\tau}_{[\alpha\beta] n}(x)=0,\qquad 
\pd^\alpha M^{\rm tw\,\tau}_{[\alpha\beta] n}(x)=0=
\pd^\beta M^{\rm tw\,\tau}_{[\alpha\beta] n}(x),
\qquad
\tau = 2,\tilde 3,
\end{align}
and
\begin{align}
\square \widehat{M}^{\rm tw3}_{\alpha n-1}(x)=0,\qquad 
\pd^\alpha \widehat{M}^{\rm tw3}_{\alpha n-1}(x)=0.
\end{align}

Let us point to the structural similarities of the operators 
(\ref{O_tw2}) and (\ref{vM_tw3}) on the one hand and 
(\ref{O_tw3}) and (\ref{M_tw2}) on the other hand. --
%
Furthermore, we observe that when organizing the various expressions such
that everywhere the partial derivatives appear to the right of the $k$th
power of $x^2$ the differential operations ensuring the symmetry type 
are transformed according to $\pd_\alpha\rightarrow {\cal D}_\alpha(k)$
whereas the generalized (first order) inner derivative 
$x_{[\alpha}\pd_{\beta]}$ remains 
unchanged because it commutes with $x^2$. This might be used as a hint how 
to extend harmonically the operators when they are given on the light-cone 
in terms of inner derivatives $\d_\alpha\equiv\tilde{\cal D}_\beta(0)$ 
and $\lcx_{[\alpha}\tilde\pd_{\beta]}$. Namely, a short look on our results, 
Eqs.~(\ref{O_tw2}), (\ref{O_tw3}), (\ref{N_tw3}), (\ref{M_tw2}), 
(\ref{M_tw3a}) and (\ref{M_tw3b}), shows that the expressions on-cone 
are obtained by restricting the sums to those terms with $k=0$ and replacing 
$x\rightarrow \lcx,\;\pd_\alpha\rightarrow{\pd}/{\pd\lcx^\alpha}$.
Let us mention that this is nothing else but another form of the 
harmonic extension which is well-known in group theory 
\cite{BT77,Dobrev77,Dobrev82}. 

In the case of symmetric traceless tensors the
harmonic extension is uniquely defined and proved in Ref~\cite{BT77}. 
In all the other 
cases it can be used heuristically according to the above observation.
This harmonic extension gives an one-to-one relation between homogeneous 
polynomials on the light-cone and the corresponding harmonic polynomials 
off-cone. It is important to note that the unique harmonic extension 
must not destroy the type of the Lorentz representation. 

According to Eqs.~(\ref{O_tw2}), (\ref{O_tw3}), (\ref{N_tw3}), 
(\ref{M_tw2}), (\ref{M_tw3a}) and (\ref{M_tw3b}) we may introduce the 
orthogonal projection operators onto the subspaces with definite spin and, 
therefore, with definite geometric twist $\tau = 2,3$ as follows:
\begin{align}
\label{OPROJ}
O^{(\tau)}_{(5)\alpha n}(x) 
&= {\cal P}^{(\tau)\,\mu}_{\alpha n} O_{(5)\mu n}(x),\\
\label{NPROJ}
N^{(\tau)}_{n}(x) &= {\cal P}^{(\tau)}_n N_n(x),\\
\label{MPROJ}
M^{(\tau)}_{[\alpha\beta]n}(x) &= 
{\cal P}^{(\tau)\,[\mu\nu]}_{[\alpha\beta]n}M_{[\mu\nu]n}(x).
\end{align}
The corresponding twist projectors onto the light-cone are discussed 
in more detail in Ref.~\cite{GL01}.

\subsection{Resummed harmonic operators with twist $\tau = 2,\,3$}

In this subsection we rewrite the harmonic operators of definite
twist $\tau = 2, 3$ which have been given in Subsect.~A 
in terms of Gegenbauer polynomials
analogous to the presentation in Sect.~II for the scalar 
harmonic operators. In a second step we resum the infinite series
(for $n$) of Gegenbauer polynomials to the related Bessel functions  
thereby obtaining nonlocal harmonic operators of (the same) definite twist.

For notational simplicity we introduce the following abbreviations
for the homogeneous polynomials and their related functions:
\begin{align}
h_n^\nu(q|x)&=
\left(\hbox{\large$\frac{1}{2}$}\sqrt{q^2 x^2}\right)^n
\!C_n^\nu\bigg(\frac{qx}{\sqrt{q^2 x^2}}\bigg),
\\
{\cal H}_\nu(q|x) &= 
\sqrt{\pi}
\left(\sqrt{(qx)^2-q^2 x^2}\right)^{1/2-\nu}
\!J_{\nu-1/2}\left(\hbox{\large$\frac{1}{2}$}
\sqrt{(qx)^2-q^2 x^2}\right)
\e^{\ii qx/2}.
\end{align}
Using the following functional relations of the Gegenbauer and
Bessel functions \cite{PBM}, 
\begin{gather*}
m C_m^\nu(z) 
= 2\nu\left(zC_{m-1}^{\nu+1}(z) - C_{m-2}^{\nu+1}(z)\right),
\\
\frac{\d}{\d z} \left( z^{-\lambda} J_\lambda(z) \right)
= - z^{-\lambda} J_{\lambda+1}(z) ,
\\
z \frac{\d}{\d z}  J_\lambda(z)
= -\lambda J_{\lambda}(z) + z J_{\lambda-1}(z),
\end{gather*}
respectively, one obtains the following results for the partial derivations
of these functions:
\begin{align}
\pd_\alpha h_n^\nu(q|x)
&= \nu \left( q_\alpha h_{n-1}^{\nu+1}(q|x)
 - \hbox{\large$\frac{1}{2}$} q^2 x_\alpha h_{n-2}^{\nu+1}(q|x) \right),
\\
\pd_\alpha {\cal H}_\nu(q|x) 
&= \left( \ii q_\alpha (2\nu+1+x\pd) 
- \hbox{\large$\frac{1}{2}$}(\ii q)^2 x_\alpha \right){\cal H}_{\nu+1}(q|x).
\end{align}

The local twist-2 vector operator is obtained through derivation with 
respect to $x^\alpha$ of the local twist-2 scalar operator as follows,
cf.,~Eq.~(\ref{O_tw2}),
\begin{align}
\label{XO2}
O^{\text{tw2}}_{\alpha n}(x) 
&=
\frac{1}{(n+1)^2}\,\pd_\alpha
\int\!\d^4 q\, \big(\bar{\psi}\gamma_\mu\psi\big)(q)
\left\{ x^\mu\, h_n^2(q|x)
-\hbox{\large$\frac{1}{2}$}\,  q^\mu x^2\, h_{n-1}^2(q|x)
\right\}
\nonumber\\
&=
\frac{1}{(n+1)^2}
\int\!\d^4 q\, \big(\bar{\psi}\gamma_\mu\psi\big)(q)
\bigg\{ 
\delta_\alpha^\mu \,h_n^2(q|x)
- q^\mu x_\alpha \,h_{n-1}^2(q|x)  
\\
&\qquad\qquad\qquad
+2 x^\mu q_\alpha \,h_{n-1}^3(q|x)
-\big(x^\mu x_\alpha q^2+q^\mu q_\alpha x^2\big) \,h_{n-2}^3(q|x)
+\hbox{\large$\frac{1}{2}$}\, q^\mu x_\alpha x^2 q^2 \,h_{n-3}^3(q|x)
\bigg\},
\nonumber
\\
\intertext{and the corresponding resummed nonlocal twist-2 vector 
operator reads}
\label{nl_O2}
O^{\rm tw2}_\alpha (x,-x)
&= \pd_\alpha \int_0^1\d t
\int\!\d^4 q\, \big(\bar{\psi}\gamma_\mu\psi\big)(q)
\left\{ x^\mu \left(2+q\pd_q\right)
-\hbox{\large$\frac{1}{2}$}\,\ii\,t  q^\mu\, x^2\right\}
\left(3+q\pd_q\right)\, {\cal H}_2(q|tx)
\nonumber\\
&=\int\d^4 q\, \big(\bar{\psi}\gamma_\mu\psi\big)(q)
\left(2+q\pd_q\right)\int_0^1\d t
\bigg\{
\Big[
\left(3+q\pd_q\right)\delta^\mu_{\alpha}
-\ii t q^\mu x_\alpha\Big]
{\cal H}_2(q|tx)
\\
&\qquad\qquad
+\Big[
\left(3+q\pd_q\right)
\left(
\left(4+q\pd_q\right)\ii t q_\alpha  x^\mu 
-\hbox{\large$\frac{1}{2}$}(\ii t)^2 \big(q^2 x^\mu x_\alpha
+ x^2 q^\mu q_\alpha \big) 
\right)
+\hbox{\large$\frac{1}{4}$}
(\ii t)^3 q^\mu q^2 x_{\alpha}x^2 \Big]
{\cal H}_3(q|tx)
\bigg\}.
\nonumber
\end{align}
Here, the equality $1/(n+1) = \int_0^1\d t\, t^n$ has been used;
analogous equalities will be used in the following.

The resummed local twist-3 vector operator reads, cf.,~Eq.~(\ref{O_tw3}):
\begin{align}
\label{XO3}
O^{\rm tw3}_{\alpha n}(x)
&=\frac{2}{n+1}\, x^\beta
\Big\{\delta^\mu_{[\alpha}\pd_{\beta]}-\frac{1}{n+1}\, 
x_{[\alpha}\pd_{\beta]}\pd^{\mu}\Big\}
\int\d^4 q\, \big(\bar{\psi}\gamma_\mu\psi\big)(q) \,h_{n}^1(q|x)
\nonumber\\
&=\frac{1}{(n+1)^2}
\int\d^4 q\, \big(\bar{\psi}\gamma_\mu\psi\big)(q) x^\beta
\bigg\{ 2(n+1)
\delta_{[\alpha}^\mu q_{\beta]}  \,h_{n-1}^2(q|x)
+(n+2)\, x_{[\alpha}\delta_{\beta]}^\mu q^2 \,h_{n-2}^2(q|x)   
\\
&\qquad\qquad\qquad\qquad\qquad\qquad\quad
-4 x_{[\alpha}q_{\beta]} q^\mu \,h_{n-2}^3(q|x) 
+2 x_{[\alpha} q_{\beta]} x^\mu q^2 \,h_{n-3}^3(q|x)
\bigg\},
\nonumber
\\
\intertext{and the bilocal twist-3 vector operator is given by}
\label{nl_O3}
O^{\rm tw3}_\alpha (x,-x)
&=2\int\d^4 q\, \big(\bar{\psi}\gamma_\mu\psi\big)(q)\,
\int_0^1\d t\,x^\beta\Big\{
\delta^\mu_{[\alpha}\pd_{\beta]}\,(1+q\pd_q)
-x_{[\alpha}\pd_{\beta]}\pd^\mu\Big\}
{\cal H}_1(q|tx)
\nonumber\\
&=
\int\d^4 q\, \big(\bar{\psi}\gamma^\mu\psi\big)(q)
\int_0^1\d t \, \left(2+q\pd_q\right)\, x^\beta
\bigg\{\Big[
2\,\ii t\,\delta^\mu_{[\alpha}q_{\beta]}(2+q\pd_q)\,
-(\ii t)^2 \delta^\mu_{[\alpha}x_{\beta]} q^2 \Big] {\cal H}_2(q|tx)
\\
&\qquad\qquad\qquad\qquad\qquad\qquad\qquad\qquad
- \ii t\, x_{[\alpha} q_{\beta]} 
\Big[2\, \ii t\, q^\mu  \left(5+q\pd_q\right)
-(\ii t)^2 x^\mu  q^2  \Big]\,{\cal H}_3(q|tx)
\bigg\}.
\nonumber
\end{align}
Let us remark that in the expression (\ref{nl_O3}) by partial 
integration the term containing $(1+t\pd_t)$ contributes only through 
its surface terms for $t=1$. Remind also that according to the special
structure of ${\cal H}_\nu(q|tx) \equiv {\cal H}_\nu(tq|x)$ in 
Eqs.~(\ref{nl_O2}) and (\ref{nl_O3}) the homogeneous derivation $q\pd_q$ 
may be replaced by either $x\pd_x$ or $t\pd_t$. Analogous observations
could be made below; however, there the $x-$derivatives in the local 
as well as the nonlocal case will not be taken explicitly.

The local twist-3 chiral-odd scalar operator will be given only
for completeness, cf.,~Eq.~(\ref{Nttw_GP}):
\begin{align}
\label{XN3}
N^{\rm tw3}_n(x)
&=
\int\d^4 q\, \big(\bar{\psi}\psi\big)(q) \,h_{n}^1(q|x),
\\
\intertext{and for the corresponding bilocal one, one gets}
\label{nl_N3}
N^{\rm tw3}(x,-x)
&=\int\d^4 q\, \big(\bar{\psi}\psi\big)(q)
\left(1+q\pd_q\right)\,{\cal H}_1(q|x).
\end{align}

The local resummed twist-2 chiral-odd skew tensor operator is given by,
cf.,~Eq.~(\ref{M_tw2}),
\begin{align}
\label{XM2}
M^{\rm tw2}_{[\alpha\beta] n}(x)
&=\frac{2}{n+2}\,
\Big\{\delta^\mu_{[\alpha}\pd_{\beta]}-\frac{1}{n+2}\, 
x_{[\alpha}\pd_{\beta]}\pd^{\mu}\Big\}
\int\d^4 q\, \big(\bar{\psi}\sigma_{\mu\nu}\psi\big)(q)
\Big\{
x^\nu \,h_{n}^2(q|x)
-\hbox{\large$\frac{1}{2}$}\,q^\nu\,x^2 \,h_{n-1}^2(q|x)
\Big\},
\\
\intertext{and the related bilocal twist-2 skew tensor operator reads}
\label{nl_M2}
M^{\mathrm tw 2}_{[\alpha\beta]} (x,-x)
&=2\int\!\d^4 q\, \big(\bar{\psi}\sigma_{\mu\nu}\psi\big)(q)
\int_0^1\!\d t\,t\,(1+q\pd_q)
\Big\{(2+q\pd_q)\,\delta^\mu_{[\alpha}\pd_{\beta]} 
-x_{[\alpha}\pd_{\beta]}\pd^\mu \Big\} 
\Big[  x^\nu \left(3+q\pd_q\right)
-\hbox{\large$\frac{1}{2}$}\,\ii\,t q^\nu x^2 \Big]
{\cal H}_2(q|tx).
\end{align}
For the related, resummed local twist-3 skew tensor operator 
resulting from the trace terms we obtain
\begin{align}
\label{XM3}
{M}^{\rm tw3}_{[\alpha\beta] n}(x)
&=\frac{2}{n(n+2)}\,x_{[\alpha} \pd_{\beta]}
\int\d^4 q\, \big(\bar{\psi}\sigma_{\mu\nu}\psi\big)(q)\, q^\mu x^\nu
\,h_{n-1}^2(q|x),
\\
\intertext{leading to the following twist-3 bilocal skew tensor}
\label{nl_M3}
M^{\rm tw3}_{[\alpha\beta]} (x,-x)
&=2 x_{[\alpha}\pd_{\beta]}
\int\d^4 q\, 
\big(\bar{\psi}\sigma_{\mu\nu}\psi\big)(q)\, \ii\, q^\mu x^\nu
(2+q\pd_q)
\int_0^1\d t\,{\cal H}_2(q|tx).
\end{align}

In addition, the independent twist-3 skew tensor having symmetry type 
$[n,1,1]$ has the following local form:
\begin{align}
\label{XtM3}
\widetilde M^{\rm tw3}_{[\alpha\beta] n}(x)
&=\frac{3}{n+2} x^\gamma
\bigg\{\delta^\mu_{[\alpha}\delta^\nu_{\beta}\pd_{\gamma]}
-\frac{2}{n}\, \delta^\mu_{[\alpha}x_{\beta}\pd_{\gamma]}\pd^{\nu}\bigg\}
\int\d^4 q\, \big(\bar{\psi}\sigma_{\mu\nu}\psi\big)(q) 
\,h_{n}^1(q|x)
\\
\intertext{leading to the following expression for 
the twist-3 bilocal skew tensor,}
\label{nl_tM3}
\widetilde M^{\rm tw3}_{[\alpha\beta]} (x,-x)
& =3 x^\gamma 
\bigg\{\delta^\mu_{[\alpha}\delta^\nu_{\beta}\pd_{\gamma]} x\pd
-  2\, \delta^\mu_{[\alpha}x_{\beta}\pd_{\gamma]}\pd^{\nu}\bigg\}
\int\d^4 q\, 
\big(\bar{\psi}\sigma_{\mu\nu}\psi\big)(q)\,
(1+q\pd_q)
\int_0^1 \d t \frac{1-t^2}{2t}\,
{\cal H}_1(q|tx).
\end{align}

\section{Power Corrections of non-forward matrix elements}
\setcounter{equation}{0}

\subsection{Double distributions from quark--antiquark operators
with definite twist $\tau$}

According to the different approaches exemplified in Subsect.~II.B and 
II.C in order to obtain the non-forward matrix elements for the 
quark--antiquark operators
with definite twist $\tau$ we have only to take matrix elements of
the expressions given in the last Subsect. IV.B and to use 
formula (\ref{NLME}) of Subsect. II.A. Let us emphasize that
the spin-dependence of the hadronic states $|P_i, S_i\rangle, i = 1,2$,
is completely contained in the independent kinematical structures
${\cal K}^a_\Gamma({\mathbb P})$ being relevant for the processes
under consideration, e.g., the Dirac and the Pauli structure 
$\bar u(P_2, S_2) \gamma_\mu u(P_1,S_1)$ and 
$\bar u(P_2, S_2) \sigma_{\mu\nu} P_-^\nu u(P_1,S_1)/M$, respectively,
in case of the virtual Compton scattering (see,~e.g.,~\cite{BGR99}).

Let us now give the final expressions for matrix elements of
those nonlocal operators 
which are relevant for the power corrections of the various
processes. The corresponding expressions for the local operators are 
easily obtained by taking moments.

For the non-forward matrix element of the nonlocal twist-2 and twist-3 
vector operator, Eqs.~(\ref{nl_O2}) and (\ref{nl_O3}), 
obeying the replacement (\ref{NLME}), one gets
\begin{align}
\langle P_2, S_2 |O^{\rm tw2}_\alpha &(\kappa x,-\kappa x) |P_1,S_1\rangle
= {\cal K}^{a}_{\mu}({\mathbb P})
\int{\mathrm D}{\mathbb Z}\, F^{(2)}_a({\mathbb Z})\,
\pd_\alpha \int_0^1\!\!\d t
\left\{ x^\mu \left(2+x\pd\right)
-\hbox{\large$\frac{1}{2}$}\,\ii\, \kappa t\,
({\mathbb P}^\mu{\mathbb Z})\, x^2\right\}
\left(3+x\pd\right)\, {\cal H}_2({\mathbb PZ}|\kappa tx)
\nonumber\\
&={\cal K}^{a}_\mu({\mathbb P})
\int{\mathrm D}{\mathbb Z}\, F^{(2)}_a({\mathbb Z})
\int_0^1\!\!\d t\left(2+x\pd\right)
\bigg\{
\Big[
\left(3+x\pd\right)\delta^\mu_{\alpha}
-\ii\,\kappa t\,({\mathbb P}^\mu{\mathbb Z})\,  x_\alpha\Big]
{\cal H}_2({\mathbb PZ}|\kappa tx)
\nonumber\\
&\qquad
+\Big[
\left(3+x\pd\right)
\left(
\left(4+x\pd\right)\ii\,\kappa t\, ({\mathbb P}_\alpha{\mathbb Z}) x^\mu 
-\hbox{\large$\frac{1}{2}$}
(\ii\kappa t)^2 \big(({\mathbb PZ})^2 x^\mu x_\alpha
+ ({\mathbb P}^\mu{\mathbb Z}) ({\mathbb P}_\alpha{\mathbb Z})\, x^2 \big) 
\right)
\nonumber\\
&\quad\qquad\qquad\qquad\quad\qquad\qquad\qquad\qquad\qquad
+\hbox{\large$\frac{1}{4}$}
(\ii \kappa t)^3\, ({\mathbb P}^\mu{\mathbb Z}) 
({\mathbb PZ})^2  x_{\alpha}x^2 \Big]
{\cal H}_3({\mathbb PZ}|\kappa tx)
\bigg\},
\\
\intertext{and}
\label{}
\langle P_2, S_2 |O^{\rm tw3}_\alpha &(\kappa x,-\kappa x)|P_1,S_1\rangle
=2{\cal K}^{a}_{\mu}({\mathbb P})
\int{\mathrm D}{\mathbb Z}\, F^{(3)}_a({\mathbb Z})\,
\int_0^1\!\!\d t\,
\,x^\beta\Big\{
\delta^\mu_{[\alpha}\pd_{\beta]}\,(1+x\pd)
-x_{[\alpha}\pd_{\beta]}\pd^\mu\Big\}
{\cal H}_1({\mathbb PZ}|\kappa tx)
\nonumber\\
&=
{\cal K}^{a}_{\mu}({\mathbb P})
\int{\mathrm D}{\mathbb Z}\, F^{(3)}_a({\mathbb Z})\,
\left(2+x\pd\right) \int_0^1\!\!\d t\, x^\beta
\bigg\{\Big[
2\,\ii\kappa t\,\delta^\mu_{[\alpha}
({\mathbb P}_{\beta]}{\mathbb Z})(2+x\pd) 
- (\ii\kappa t)^2 \delta^\mu_{[\alpha}x_{\beta]} ({\mathbb PZ})^2 \Big] 
{\cal H}_2({\mathbb PZ}|\kappa tx)
\nonumber\\
&\qquad\qquad\qquad\qquad\qquad\qquad
-\ii\kappa t\, x_{[\alpha}({\mathbb P}_{\beta]}{\mathbb Z}) 
\Big[2 \ii\kappa t\,({\mathbb P}^\mu{\mathbb Z}) \left(5+x\pd\right)
- (\ii\kappa t)^2 x^\mu ({\mathbb PZ})^2 
\Big]\,{\cal H}_3({\mathbb PZ}|\kappa tx)
\bigg\}.
\end{align}

In the case of the axial vector operators 
$O^{\tau}_{5\alpha}(\kappa x,-\kappa x)$ the kinematical structures
${\cal K}_{5\mu}({\mathbb P})$ contain an additional $\gamma_5$ and the 
double distributions should be denoted by $G^{(\tau)}_{a}(\mathbb Z)$ 
but otherwise the structure will be exactly the same.

The non-forward matrix elements of the 
nonlocal twist-3 chiral-odd scalar operator, Eq.~(\ref{nl_N3}), reads:
\begin{align}
\label{}
\langle P_2,& S_2 |
N^{\rm tw3}(\kappa x,-\kappa x) |P_1,S_1\rangle
={\cal K}^{a}({\mathbb P})
\int{\mathrm D}{\mathbb Z}\, E^{(3)}_a({\mathbb Z})\,
\left(1+x\pd\right)\,{\cal H}_1({\mathbb PZ}|\kappa x).
\end{align}

The non-forward matrix elements
nonlocal twist-2 chiral-odd skew tensor operator and of the related 
twist-3 operator, Eqs.~(\ref{nl_M2}) and (\ref{nl_M3}), are given by
\begin{align}
\label{}
\langle P_2, S_2 |
M^{\mathrm tw2}_{[\alpha\beta]}(\kappa x,-\kappa x)|P_1,S_1\rangle
&=2{\cal K}^{a}_{[\mu\nu]}({\mathbb P})
\int{\mathrm D}{\mathbb Z}\, H^{(2)}_a({\mathbb Z})\,
\int_0^1\!\!\d t\,t\, (1+x\pd)
\Big\{(2+x\pd) \delta^\mu_{[\alpha}\pd_{\beta]}
-x_{[\alpha}\pd_{\beta]}\pd^\mu \Big\}
\nonumber\\
&\quad\qquad\qquad\qquad\qquad\qquad
\times
\Big[  x^\nu \left(3+x\pd\right)
-\hbox{\large$\frac{1}{2}$}\,\ii\kappa t\, 
({\mathbb P}^\nu{\mathbb Z}) x^2 \Big]
{\cal H}_2({\mathbb PZ}|\kappa tx),
\\
\intertext{and}
\label{}
\langle P_2, S_2 |
M^{\mathrm tw3}_{[\alpha\beta]}(\kappa x,-\kappa x)|P_1,S_1\rangle
&=2 {\cal K}^{a}_{[\mu\nu]}({\mathbb P})
\int{\mathrm D}{\mathbb Z}\, H^{(3)}_a({\mathbb Z})\,
x_{[\alpha}\pd_{\beta]}\,
 \ii\kappa \, ({\mathbb P}^\mu{\mathbb Z}) x^\nu (2+x\pd)
\int_0^1\!\!\d t\,{\cal H}_2({\mathbb PZ}|\kappa tx).
\end{align}

Finally, the non-forward matrix element of the
independent nonlocal twist-3 skew tensor, Eq.~(\ref{nl_tM3}), reads:
\begin{align}
\label{}
\langle P_2, S_2 |
\widetilde M^{\mathrm tw3}_{[\alpha\beta]}(\kappa x,-\kappa x)
|P_1,S_1\rangle
&={3} \widetilde{\cal K}^{a}_{[\mu\nu]}({\mathbb P})\!
\int\!{\mathrm D}{\mathbb Z}\, \widetilde H^{(3)}_a({\mathbb Z})\,
x^\gamma 
\Big\{\delta^\mu_{[\alpha}\delta^\nu_{\beta}\pd_{\gamma]} x\pd
-  2\, \delta^\mu_{[\alpha}x_{\beta}\pd_{\gamma]}\pd^{\nu}\Big\}
\nonumber\\
&\hspace{5cm}
\times(1+x\pd)\int_0^1\!\!\d t\, \frac{1\!-\!t^2}{2t}\,
{\cal H}_1({\mathbb PZ}|\kappa tx).
\end{align}

The matrix elements of the scalar and vector operators are obtained
by multiplying the above expressions by $x^\alpha$ (of course, for the 
operator $M^{\mathrm tw3b}_{[\alpha\beta]}(\kappa x,-\kappa x)$ 
no vector operator exists because of its symmetry type). For completeness 
let us also give the
power corrections of the double distributions of higher twist in the
scalar cases. Using the expressions (\ref{O^(2+2j)}), (\ref{N^(3+2j)})
and (\ref{M^(2+2j)}) we obtain
\begin{align}
&\langle P_2, S_2 |O^{{\rm tw}(2+2j)}(\kappa x,-\kappa x)|P_1,S_1\rangle
= {\cal K}^{a}_{\mu}({\mathbb P})
\int{\mathrm D}{\mathbb Z}\, F^{(2+2j)}_a({\mathbb Z})\,
(\ii {\mathbb PZ})^{2j} 
\\
&\hspace{3cm}
\times 
\bigg\{\Big[ x^\mu \left(2+x\pd\right)
-\hbox{\large$\frac{1}{2}$}\,\ii\, \kappa \,
({\mathbb P}^\mu{\mathbb Z})\, x^2\Big]
\left(3+x\pd\right)\, {\cal H}_2({\mathbb PZ}|\kappa x)
-2j \frac{\ii{\mathbb P}_\mu {\mathbb Z}}{({\mathbb PZ})^2}
\left(1+x\pd\right) {\cal H}_1({\mathbb PZ}|\kappa x)\bigg\},
\nonumber\\
&\langle P_2, S_2 |N^{{\rm tw}(3+2j)}(\kappa x,-\kappa x)|P_1,S_1\rangle
= {\cal K}^{a}({\mathbb P})
\int{\mathrm D}{\mathbb Z}\, E^{(3+2j)}_a({\mathbb Z})\,
(\ii {\mathbb PZ})^{2j}
\left(1+x\pd\right) {\cal H}_1({\mathbb PZ}|\kappa x)
\\
\intertext{and}
&\langle P_2, S_2 |M^{{\rm tw}(3+2j)}(\kappa x,-\kappa x)|P_1,S_1\rangle
= {\cal K}^{a}_{[\mu\nu]}({\mathbb P})
\int{\mathrm D}{\mathbb Z}\, H^{(3+2j)}_a({\mathbb Z})\,
\ii{\mathbb P}^\mu{\mathbb Z} x^\nu (\ii {\mathbb PZ})^{2j}
\left(2+x\pd\right)\left(3+x\pd\right)\, {\cal H}_2({\mathbb PZ}|\kappa x),
\end{align}
respectively.
Let us remind that, eventually, the $t$-integration which appeared, e.g.,
in Eq.~(\ref{N_twn_sc}) should be included 
into the definition of the operators of definite twist and, therefore,
it could appear also here.

Now, a couple of remarks are in order. First, as indicated, different
tensorial structures of the operators lead to different kinematical 
structures. In the
forward case they simplify or eventually disappear, e.g., the Dirac
structures are to be replaced by $P_\mu$ and $S_\mu$ for the vector
and axial vector case, respectively, whereas the Pauli structures
vanish. Second, going on-cone the double distributions remain the same.
Therefore, their evolution is determined by the anomalous dimensions
resulting from the renormalization group equation of the corresponding 
light-ray operators. Third, taking the forward limit on the light-cone the
double distributions are projected onto the quark distribution functions
determined in Ref.~\cite{GL01}. In this connection we emphasize that
these distribution functions are uniquely determined by the geometric twist
which differs from the dynamic twist being used for phenomenological
considerations. 

Furthermore, we have to mention that -- contrary to the case when 
the non-forward matrix elements are restricted to the light-cone where
the decomposition of the non-local quark-antiquark operators into 
operators of definite twist terminates at finite values $\tau_{\rm max}$ --
the off-cone decomposition results in an infinite series of any twist. 
Therefore,
the decomposition of the double distributions related to the undecomposed
operators, $f_a({\mathbb Z}, {\mathbb P}_i {\mathbb P}_j, x^2, \mu^2)$, 
results also in an infinite series. This becomes obvious especially in the
scalar cases considered in Sec.~III.
 
The double distributions occur in
the scattering amplitudes of various physical processes, like virtual
Compton scattering in the generalized Bjorken region. There, after 
multiplication with appropriate coefficient functions $C_\Gamma(x)$
a Fourier transformation with respect to $x$ has to performed relating
$x$ to the (inverse) momentum transfer $1/Q$ of the process. Thereby,
the variables $(x{\mathbb PZ})$ and $x^2 ({\mathbb PZ})^2$ are changed
into $(Q{\mathbb PZ})/Q^2$ and $({\mathbb PZ})^2/Q^2$. The Fourier
transformation will be simplified when the Poisson integral is used
for the Bessel functions. This will be considered in more detail
in another paper.

Another application of our general procedure is the consideration
of the meson distribution functions which does not require such an
additional Fourier transformation.

\subsection{Mass corrections of vector meson distributions}

Based on bilocal light-cone operators of definite twist-2, 3 and 4 the 
distribution amplitudes (DAs) of the $\rho$-meson have been already studied 
in Refs.~\cite{L01a,L01b}. The relationship between the DAs
of geometric and dynamical twist has also been discussed 
(see, also, Ref.~\cite{BL01}).
These considerations can be extended to the other vector mesons by trivial 
substitutions. It is an advantage of this approach that operators of 
well-defined twist are used and, therefore, it is not necessary to use 
operator relations, like dynamical Wandzura-Wilczek relations, 
in order to isolate contributions of twist-3 and 4. 
The difference, in the non-forward case, between geometric and dynamical
Wandzura-Wilczek relations was pointed out in Refs.~\cite{L01b,BL01}.

In this Subsection we extend that analysis of the $\rho$-meson DAs
to include all meson momentum resp. mass terms of harmonic 
twist-2 and 
twist-3 off-cone operators in order to get the power corrections to
the results obtained earlier. Vector meson mass corrections of order $x^2$ 
were already discussed by Ball and Braun~\cite{BB99} and resummed 
meson mass corrections in the scalar case was given by Ball~\cite{Ball99} 
(see also Ref.~\cite{BB91}).

The structure of the mass corrections of $\rho$-meson DAs 
of geometric twist are, from the group theoretical point of view, 
similar to the target mass corrections in deep inelastic scattering,
which can be resummed using Nachtmann's method~\cite{Nachtmann73}.
Also here, the basic tool, in order to obtain the mass corrections, 
are the harmonic operators of definite geometric twist which have been
determined in Sec.~IV. Obviously, they are the harmonic extensions 
of the corresponding light-cone operators which already are used in 
Refs.~\cite{L01a,L01b} for the classification of the
corresponding meson light cone DAs with respect to geometric twist.
 
Let us now introduce the distribution functions for the harmonic operators 
 of geometric twist sandwiched between the vacuum and the meson state,
$\langle 0|O^{(\tau)}_{\Gamma}(x,-x)|\rho(P,\lambda)\rangle $. Thereby,
we adopt the definitions of Chernyak and Zhitnitsky \cite{CZ84} in the
terminology of Ref.~\cite{BBKT98}.
As usual, these matrix elements are related to 
the momentum $P_\alpha$ and polarization vector $e^{(\lambda)}_\alpha$ 
of the meson with helicity $\lambda,\,P^2=m^2$, 
$e^{(\lambda)}\cdot e^{(\lambda)}=-1$, 
$P\cdot e^{(\lambda)}=0$, $m$ denotes the meson mass.

First, we consider the chiral-even {\em  vector operator}. Using the 
twist projections we introduce the moments of the meson DAs,
$\Phi^{(\tau)}_n$, of twist $\tau$ according to
\begin{align}
\label{}
\langle 0|O^{(\tau)}_{\alpha n}(x)|\rho(P,\lambda)\rangle 
= f_\rho m\,{\cal P}^{(\tau)\beta}_{\alpha n}
\left(e^{(\lambda)}_\beta (Px)^n \Phi^{(\tau)}_n\right),
\end{align} 
where $f_\rho$ is the vector meson decay constant.
The corresponding meson distribution amplitudes $\hat\Phi^{(\tau)}(\xi)$ 
are given by inverting the moment integral,
\begin{align}
\label{WF}
\Phi^{(\tau)}_n=\int_{-1}^1\d \xi\, \xi^n \hat\Phi^{(\tau)}(\xi),\quad
\xi=u-(1-u)=2u-1.
\end{align} 
These distribution amplitudes are dimensionsless functions of $\xi$ and 
describe the probability
amplitudes to find the $\rho$-meson in a state with minimal number of 
constituents (quark and antiquark) which carry the momentum fractions 
$u$ (quark) and $(1-u)$ (antiquark).

Taking into account expression~(\ref{XO2}) we obtain the local  
twist-2 matrix element as follows:
\begin{align}
\label{lmatrix_O_tw2}
\langle 0|O^{\rm tw2}_{\alpha n}(x)&|\rho(P,\lambda)\rangle
=\frac{1}{(n+1)^2}\,f_\rho m\,\Phi_n^{(2)}
\Big\{
e^{(\lambda)}_\alpha h^2_n(P|x)
+2P_\alpha(e^{(\lambda)}x) h^3_{n-1}(P|x)
-m^2 x_\alpha (e^{(\lambda)}x)h^3_{n-2}(P|x)
\Big\},
\\
\intertext{which is analogous to Wandzura's expression~\cite{Wandzura77} 
for the target mass corrections in deep inelastic scattering in the 
$x$-space. After resummation  
we get the bilocal matrix element of geometric twist-2:}
\label{nmatrix_O_tw2}
\langle 0|O^{\rm tw2}_{\alpha}(x,&-x)|\rho(P,\lambda)\rangle 
=f_\rho m\int_0^1\d t\int_{-1}^1\d \xi\, \hat\Phi^{(2)}(\xi)
\left(3+\xi\pd_\xi\right)\left(2+\xi\pd_\xi\right)
\nonumber\\
&\qquad\qquad\qquad\qquad\times
\bigg\{e^{(\lambda)}_\alpha
{\cal H}_2(P\xi| t x)
+\bigg(\left(4+\xi\pd_\xi\right)(\ii\xi t) P_\alpha
-\frac{1}{2}(\ii\xi t)^2 m^2 x_\alpha\bigg) 
(e^{(\lambda)}x){\cal H}_3(P\xi| t x)\bigg\}.
\\
\intertext{An analogous calculation using expression (\ref{XO3})
gives the local twist-3 matrix element}
\label{lmatrix_O_tw3}
\langle 0|O^{\rm tw3}_{\alpha n}(x)&|\rho(P,\lambda)\rangle
=\frac{1}{(n+1)^2}\,f_\rho m\,\Phi_n^{(3)} x^\beta
\Big\{2(n+1)\,
e^{(\lambda)}_{[\alpha} P_{\beta]} h^2_{n-1}(P|x)
+(n+2)\,m^2 x_{[\alpha}e_{\beta]}^{(\lambda)} h^2_{n-2}(P|x)
\nonumber\\
&\qquad\qquad\qquad\qquad\qquad\qquad
+2\,m^2 x_{[\alpha} P_{\beta]} (e^{(\lambda)}x) h^3_{n-3}(P|x)
\Big\};
\\
\intertext{the resummed bilocal matrix element of twist-3 reads}
\label{nmatrix_O_tw3}
\langle 0|O^{\rm tw3}_{\alpha}(x,&-x)|\rho(P,\lambda)\rangle 
=f_\rho m\int_{-1}^1\d\xi\, \hat\Phi^{(3)}(\xi)
\left(2+\xi\pd_\xi\right) x^\beta
\bigg\{2(\ii\xi) e^{(\lambda)}_{[\alpha}P_{\beta]}{\cal H}_2(P\xi| x)
\nonumber\\
&\qquad\qquad\qquad
+m^2\int_0^1\d t\Big(
(\ii\xi t)^2  x_{[\alpha} e^{(\lambda)}_{\beta]}{\cal H}_2(P\xi| t x)
+(\ii\xi t)^3 x_{[\alpha}P_{\beta]} (e^{(\lambda)}x){\cal H}_3(P\xi| t x)
\Big)\bigg\}.
\end{align}
It is well-known that the twist-3 operator $O^{\rm tw3}_{\alpha}(x,-x)$ 
is related to other twist-3 operators containing total derivatives and 
operators of Shuryak-Vainshtein type by means of QCD equations of 
motion~\cite{BB88,BB99}. Therefore, the matrix 
elements~(\ref{lmatrix_O_tw3}) and (\ref{nmatrix_O_tw3}) include the 
contribution of the twist-3 operator containing total derivatives 
which is as large as those from twist-2 (see also~\cite{BL01}). 
In addition the next higher twist contributions of the chiral-even vector 
operator are of twist-4.

Now we consider the chiral-even {\em axial vector operator }
$O_{5\alpha n}(x)$ 
with the same twist projectors as for the chiral-even vector operator 
$O_{\alpha n}(x)$.
We define the corresponding moments of the
meson DAs $\Xi^{(\tau)}_n$ by
\begin{align}
\label{}
\langle 0|O^{(\tau)}_{5\alpha n}(x)|\rho(P,\lambda)\rangle 
= \frac{1}{2}\Big(f_\rho -f_\rho^{\rm T}\frac{m_u+m_d}{m}\Big)m\,
{\cal P}^{(\tau)\beta}_{\alpha n}
\left(\epsilon_\beta^{\ \,\gamma\mu\nu}
e^{(\lambda)}_\gamma P_\mu x_\nu (Px)^n \Xi^{(\tau)}_n\right),
\end{align} 
where $f^{\rm T}_\rho$ denotes the tensor decay constant. First, 
we observe that the twist-2 contribution vanishes. The nontrivial local 
vacuum-to-meson matrix elements of this  axial vector operator 
are of twist-3:
\begin{align}
\label{lmatrix_O5_tw3}
\langle 0|O^{\rm tw3}_{5\alpha n}(x)|\rho(P,\lambda)\rangle 
&=\frac{1}{2}\Big(f_\rho-f_\rho^{\rm T}\frac{m_u+m_d}{m}\Big)m\,
\epsilon_\alpha^{\ \,\beta\mu\nu}
e^{(\lambda)}_\beta P_\mu x_\nu\,\Xi^{(3)}_n h^1_n(P|x),
\\
\intertext{and the bilocal matrix element of twist-3 reads}
\label{nmatrix_O5_tw3}
\langle 0|O^{\rm tw3}_{5\alpha}(x,-x)|\rho(P,\lambda)\rangle 
&=\frac{1}{2}
\Big(f_\rho-f_\rho^{\rm T}\frac{m_u+m_d}{m}\Big)m\,
\epsilon_\alpha^{\ \,\beta\mu\nu}
e^{(\lambda)}_\beta P_\mu x_\nu
\int_{-1}^1\d\xi\, \hat\Xi^{(3)}(\xi)
\left(1+\xi\pd_\xi\right){\cal H}_1(P\xi| x).
\end{align}
By the way the twist-4 matrix element also vanishes and the next higher 
twist contribution would be of twist-5. Also here, the matrix 
elements~(\ref{lmatrix_O5_tw3}) and (\ref{nmatrix_O5_tw3}) include the 
contribution of the twist-3 operator containing total derivatives.

The matrix element of the chiral-odd {\em scalar operator} is defined as
\begin{align}
\label{}
\langle 0|N^{\rm tw3}_{n}(x)|\rho(P,\lambda)\rangle 
=-\ii\Big( f^{\rm T}_\rho-f_\rho \frac{m_u+m_d}{m}\Big)
m^2 {\cal P}^{(3)}_{n}
\left(\big(e^{(\lambda)} x\big) (Px)^n \Upsilon^{(3)}_n\right),
\end{align}
where $\Upsilon^{(3)}_n$ is the moment of a spin-independent twist-3 
distribution function. Using expression (\ref{XN3}) the local matrix 
element is given as
\begin{align}
\label{lmatrix_N_tw3}
\langle 0|N^{\rm tw3}_{n}(x)|\rho(P,\lambda)\rangle 
&=-\ii\Big( f^{\rm T}_\rho-f_\rho \frac{m_u+m_d}{m}\Big)
\big(e^{(\lambda)} x\big) m^2
\,\Upsilon^{(3)}_n  h^1_n(P|x),
\\
\intertext{and for the bilocal matrix element of twist-3 we obtain}
\label{nmatrix_N_tw3}
\langle 0|N^{\rm tw3}(x,-x)|\rho(P,\lambda)\rangle 
&=-\ii\Big( f^{\rm T}_\rho-f_\rho \frac{m_u+m_d}{m}\Big)
\big(e^{(\lambda)} x\big) m^2
\int_{-1}^1\d \xi\, \hat\Upsilon^{(3)}(\xi)
\left(1+\xi\pd_\xi\right){\cal H}_1(P\xi| x).
\end{align}
The next higher twist contributions of the scalar chiral-odd operator
are of order twist-5.

Now, we consider the matrix elements of the chiral-odd {\em  skew tensor 
operators}. The corresponding moments of the wave function are 
introduced by
\begin{align}
\label{}
\langle 0|M^{(\tau)}_{[\alpha\beta]n}(x)|\rho(P,\lambda)\rangle 
=\ii f^{\rm T}_\rho {\cal P}^{(\tau)\mu\nu}_{[\alpha\beta]n}
\left(\Big(e^{(\lambda)}_\mu P_\nu-e^{(\lambda)}_\nu P_\mu\Big) 
(Px)^n \Psi^{(\tau)}_n\right).
\end{align}
The local matrix element of the skew tensor operator of twist-2, 
performing the differentiations in the expression (\ref{XM2}), is given by
\begin{align}
\label{lmatrix_M_tw2}
\langle 0|M^{\rm tw2}_{[\alpha\beta] n}&(x)|\rho(P,\lambda)\rangle
=\ii f^{\rm T}_\rho\,\Psi_n^{(2)} 
\bigg\{
\frac{2}{n+1}\,
e^{(\lambda)}_{[\alpha}P_{\beta]} h^2_n(P|x)
+\frac{n+3}{(n+2)^2}\,m^2 x_{[\alpha} e^{(\lambda)}_{\beta]} h^2_{n-1}(P|x)
\nonumber\\
&\qquad\qquad+\frac{m^2}{(n+2)^2(n+1)}
\Big\{
4 x_{[\alpha} e^{(\lambda)}_{\beta]} h^3_{n-1}(P|x)
-\Big(2 x_{[\alpha} e^{(\lambda)}_{\beta]}(Px)
+4x_{[\alpha}P_{\beta]} (e^{(\lambda)}x)\Big)h^3_{n-2}(P|x)
\nonumber\\
&\qquad\qquad+12x_{[\alpha}P_{\beta]} (e^{(\lambda)}x)h^4_{n-2}(P|x)
-6x_{[\alpha}P_{\beta]} (e^{(\lambda)}x)(Px) h^4_{n-3}(P|x)
\Big\}\bigg\},
\\
\intertext{and the resummed bilocal matrix element of twist-2 reads}
\label{nmatrix_M_tw2}
\langle 0|M^{\rm tw2}_{[\alpha\beta]}&(x,-x)|\rho(P,\lambda)\rangle 
=\ii\,f^{\rm T}_\rho \int_{-1}^1\d\xi \, \hat\Psi^{(2)}(\xi)
\left(3+\xi\pd_\xi\right)
\bigg\{2\left(2+\xi\pd_\xi\right) 
e^{(\lambda)}_{[\alpha}P_{\beta]}{\cal H}_2(P\xi| x)\nonumber\\
&\quad+m^2\int_0^1\d t\, t\bigg[
\left(1+\xi\pd_\xi\right)
(\ii\xi t)  x_{[\alpha} e^{(\lambda)}_{\beta]}{\cal H}_2(P\xi| t x)
\nonumber\\
&\quad
+\Big(2\left(4+\xi\pd_\xi\right)
(\ii\xi t) x_{[\alpha}e^{(\lambda)}_{\beta]} 
-(\ii\xi t)^2\big( x_{[\alpha}e^{(\lambda)}_{\beta]}(Px)
+2 x_{[\alpha}P_{\beta]}(e^{(\lambda)}x)\big)\Big){\cal H}_3(P\xi| tx)
\nonumber\\
&\quad
+\Big(2\left(5+\xi\pd_\xi\right)\left(4+\xi\pd_\xi\right)
(\ii\xi t)^2 x_{[\alpha}P_{\beta]}(e^{(\lambda)}x) 
-\left(4+\xi\pd_\xi\right)
(\ii\xi t)^3 x_{[\alpha}P_{\beta]}(e^{(\lambda)}x)(Px) \Big)
{\cal H}_4(P\xi| tx)
\bigg]\bigg\}.
\end{align}
The local matrix element of the skew tensor operator of 
twist-3  is obtained from (\ref{XM3}) as follows:
\begin{align}
\label{lmatrix_M_tw3}
\langle 0|M^{\rm tw3}_{[\alpha\beta] n}(x)|\rho(P,\lambda)\rangle
&=-\frac{2}{(n+2)n}\,\ii f^{\rm T}_\rho\,\Psi_n^{(3)} m^2
\Big\{
x_{[\alpha}e^{(\lambda)}_{\beta]} h^2_{n-1}(P|x)
+2 x_{[\alpha} P_{\beta]} (e^{(\lambda)}x) h^3_{n-2}(P|x)
\Big\},
\\
\intertext{and the corresponding bilocal matrix element reads}
\label{nmatrix_M_tw3}
\langle 0|M^{\rm tw3}_{[\alpha\beta]}(x,-x)|\rho(P,\lambda)\rangle 
&=-2\ii f^{\rm T}_\rho m^2\int_{0}^1\frac{\d t}{t} \int_{-1}^1\d\xi\, \hat\Psi^{(3)}(\xi)
\left(1+\xi\pd_\xi\right)\nonumber\\
&\quad\times\Big\{(\ii\xi t) x_{[\alpha}e^{(\lambda)}_{\beta]}{\cal H}_2(P\xi| t x)
+\left(3+\xi\pd_\xi\right)(\ii\xi t)^2  
x_{[\alpha}P_{\beta]}(e^{(\lambda)}x){\cal H}_3(P\xi| t x)\Big\}.
\end{align}
The next higher twist contributions of the skew tensor operator would
be of twist-4.
Again, the matrix elements~(\ref{lmatrix_N_tw3}), (\ref{nmatrix_N_tw3}),
(\ref{lmatrix_M_tw3}) and (\ref{nmatrix_M_tw3})
include the contribution of the twist-3 operator containing 
total derivatives.
Let us note, that the contribution of the additional twist-3 operator,
Eq.~(\ref{XtM3}), vanishes.

Finally, we remark that after projection onto the light-cone the 
matrix elements, which have been introduced in Ref.~\cite{L01a}, 
are recovered. This can be easily checked by using the expansion of the 
Gegenbauer polynomials or the Poisson integral representation of the
Bessel functions. Let us also point to the essential fact that 
the $x^2-$dependence is contained in the polynomials $h^\nu_n(P|x)$ and the 
functions
${\cal H}_\nu(P|x)$ in the case of the moments and the wave
functions, respectively. Since the latter is given by
\begin{align}
2\Gamma(\nu){\cal H}_\nu(P|x)
= 4^\nu \int_{-1}^{+1}\d t (1-t^2)^{\nu-1} 
\e^{\ii (Px) \big( 1 + t \sqrt{1-m^2 x^2/(Px)^2}\big)/2}
\end{align}
one would obtain a fairly simple expansion in powers of $m^2 x^2/(Px)^2$.
As has been pointed out for general distribution amplitudes in Sec.~II
the meson DAs of definite twist are independent from any 
coordinates or momenta and therefore, coincide on-cone 
and off-cone. The whole information about the power corrections is
contained in twist projection of the operators from which the matrix 
elements are taken. 

\section{Conclusions and Comments}

In this paper, we introduced a general procedure of parametrizing
non-forward matrix elements of non-local QCD operators
${\cal O}_\Gamma(\kappa x,-\kappa x)$ by multi-variable distribution 
amplitudes $f_a^{(\tau)}({\mathbb Z})$ of well-defined geometric twist $\tau$, 
namely, single variable hadron distributions, double distributions,
triple distributions etc., times position-dependent coefficient functions 
$c^{(\tau)}$ as well as kinematical factors ${\cal K}_\Gamma$ related to the 
Lorentz structure of the QCD operators and the hadron states 
sandwiching the operators ${\cal O}_\Gamma^{(\tau)}$.
The procedure relies on the unique twist decomposition of non-local
operators off the light-cone leading to an infinite series of 
operators with growing twist. This decomposition is completely of
group theoretical origin and is equivalent to the decomposition of
the local operators into irreducible tensor representations of the 
Lorentz group. In the case of operators whose moments are related to 
totally symmetric tensors, using the procedure of harmonic extension
\cite{BT77}, we were able to determine the decomposition completely.
For operators with non-trivial symmetry type, due to the lack of a
general theoretical framework, only the terms up to twist $\tau =3$ 
have been determined. However, this covers already almost all the 
phenomenological interesting cases of quark-antiquark operators.

Using these results we determined the off-cone power corrections
to various double distributions and the vector meson wave functions
being the inputs of the corresponding scattering amplitudes and
hadronic form factors, respectively. These power corrections, in the 
case of local operators, are expressed by a finite sum in terms of 
Gegenbauer polynomials being multiplied with the moments of the 
distribution amplitudes and, in the case of non-local operators,
by a finite sum in terms of Bessel functions now multiplied with the 
distribution amplitudes directly. As a remarkable fact, which may be
read off from Sec.~V, we observe that (the moments of) the distribution 
amplitudes are independent of whether the operators are taken off-cone 
or on-cone. Furthermore, the off-cone
expressions are obtained from the on-cone ones by harmonic extension.

This very encouraging behaviour strongly relies on the definition 
of the distribution amplitudes through matrix elements of operators 
with definite geometric twist -- contrary to their possible definition 
with respect to `dynamical' twist. This, of course, would extend a procedure 
to the non-forward case which has been introduced for the forward
matrix elements (on the light-cone) in Ref.~\cite{GL01}.
As a consequence it is not necessary 
to use any further dynamical input, like equations of motion, and to
take into account operators containing total derivatives in order
to get expressions of well-defined geometric twist: their contribution
is already contained in the expressions of given (geometric) twist.

Concerning the computation of the scattering amplitudes resp.~form factors 
of physical relevance some Fourier transformation has to be carried out
whose result mainly depends on the (singular) coefficient functions as
well as on the various Bessel functions. However, modulo specific
tensorial structures, one simply has formally to substitute 
$x\rightarrow\ii q/Q^2$ leading to the replacements of 
$(x{\mathbb PZ}) \rightarrow (q{\mathbb PZ})/Q^2$
and $ x^2 ({\mathbb PZ})^2/(x{\mathbb PZ})^2 \rightarrow
\big(({\mathbb PZ})^2/Q^2\big) \big(Q^2/(q{\mathbb PZ})\big)^2$,
especially in the expressions $h^\nu_n({\mathbb PZ}|x)$ and
${\cal H}_\nu({\mathbb PZ}|x)$. The currently most interesting 
example is the (deeply) virtual Compton scattering whose consideration
has been postponed to another paper.

There is a further point deserving attention. Since the distribution
amplitudes of definite twist are given already on-cone their 
renormalization group induced $Q^2-$evolution, which is determined
by the (non-local) anomalous dimension of the non-local light-ray 
operators of definite twist, should be clearly separated from the 
kinematical power-like target momentum resp. mass corrections.


\acknowledgments

The authors are grateful to J.~Bl\"umlein, J.~Eilers and C.~Weiss 
for various discussions. In addition, M.L. is grateful to P.~Ball  
for stimulating discussions.
He also acknowledges the Graduate College
``Quantum field theory'' at Center for Theoretical Sciences of
Leipzig University for financial support.

\end{document}